\documentclass[longauth]{aa}
\usepackage{graphicx}
\usepackage{txfonts}
\usepackage[colorlinks=true,citecolor=blue]{hyperref}
\usepackage{newtxtext,newtxmath}
\usepackage[T1]{fontenc}
\usepackage{amsmath}
\usepackage{amssymb}
\usepackage{gensymb}
\usepackage{color}
\usepackage{hyperref}
\usepackage{pdflscape}
\newcommand{\orcid}[1]{\href{https://orcid.org/#1}{\includegraphics[width=10pt]{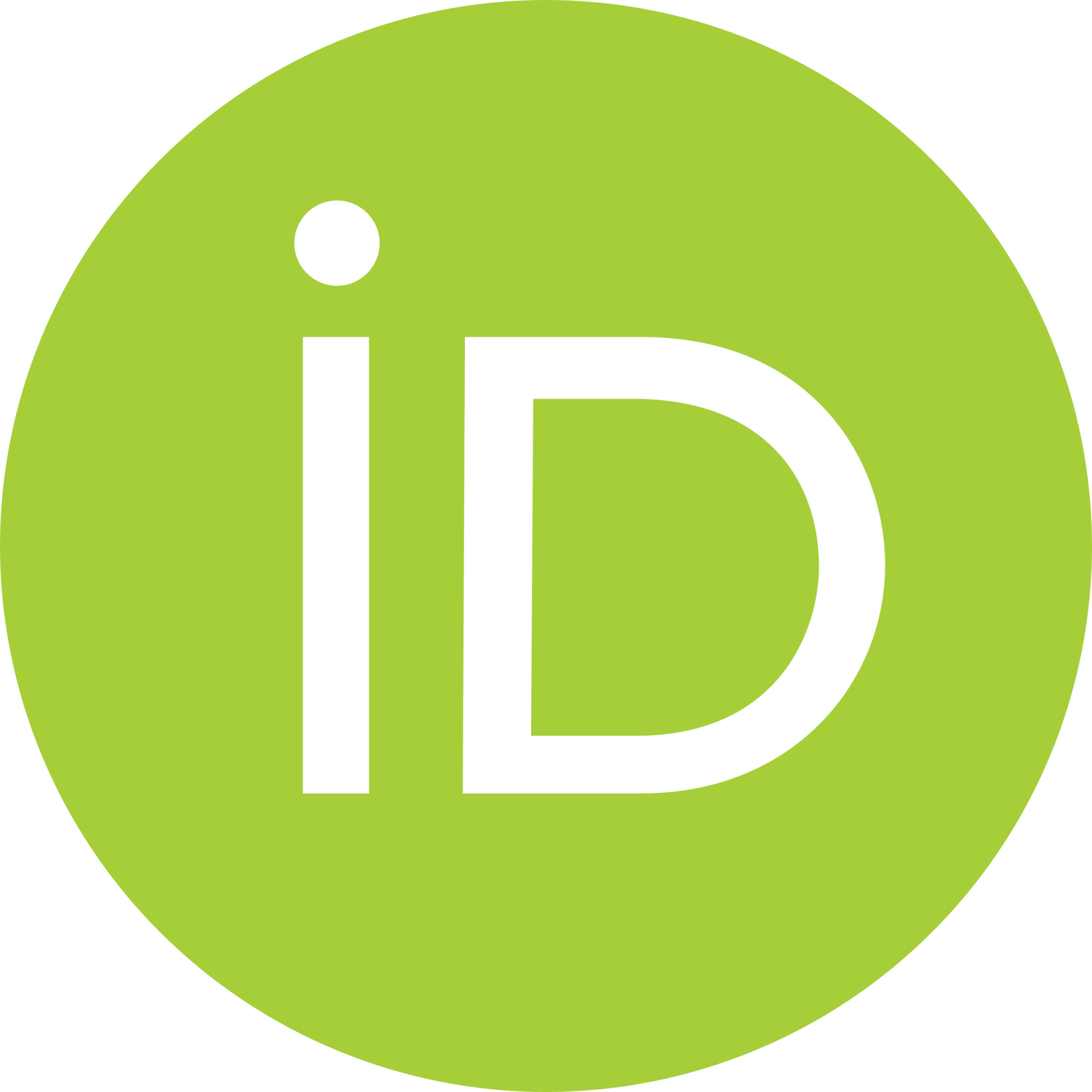}}}
\usepackage{float}
\usepackage{tabularx}
\usepackage[font=footnotesize]{caption}
\usepackage{multirow}
\usepackage{booktabs}  
\usepackage{array}      
\usepackage{siunitx}    
\usepackage[table]{xcolor}
\usepackage{colortbl}

\usepackage{appendix}

\begin{document} 

\title{Systematic uncertainties on DESI Tully-Fisher distances constrained with Integral Field Spectroscopy}
\titlerunning{DESI T-F systematics with IFS}
\authorrunning{Siwakoti et al.}

\author{Utsav~Siwakoti\inst{1,2,3}%
\thanks{\email{siwakotiutsav@ku.edu}}%
\orcid{0000-0003-3633-0098}\and
Llu\'is~Galbany\inst{1,4}\orcid{0000-0002-1296-6887}\and
Kelly~A.~Douglass\inst{5}\orcid{0000-0002-9540-546X}\and
Ariel~J.~Amsellem\inst{6}\orcid{0000-0003-3433-2698}\and
Benjamin~A.~Weaver\inst{7}\and
J.~N. Aguilar\inst{8}\and
S.~Ahlen\inst{9}\orcid{0000-0001-6098-7247}\and
D.~Bianchi\inst{10,11}\orcid{0000-0001-9712-0006}\and
D.~Brooks\inst{12}\and
T.~Claybaugh\inst{8}\and
A.~de~la~Macorra\inst{13}\orcid{0000-0002-1769-1640}\and
P.~Doel\inst{12}\and
S.~Ferraro\inst{8,14}\orcid{0000-0003-4992-7854}\and
J.~E.~Forero-Romero\inst{15,16}\orcid{0000-0002-2890-3725}\and
E.~Gazta\~naga\inst{4,17,1}\orcid{0000-0001-9632-0815}\and
S.~Gontcho~A~Gontcho\inst{18}\orcid{0000-0003-3142-233X}\and
G.~Gutierrez\inst{19}\and
C.~Hahn\inst{20}\orcid{0000-0003-1197-0902}\and
K.~Honscheid\inst{21,22,23}\orcid{0000-0002-6550-2023}\and
C.~Howlett\inst{24}\orcid{0000-0002-1081-9410}\and
D.~Huterer\inst{25,26}\orcid{0000-0001-6558-0112}\and
R.~Joyce\inst{7}\orcid{0000-0003-0201-5241}\and
S.~Juneau\inst{7}\orcid{0000-0002-0000-2394}\and
R.~Kehoe\inst{27}\and
T.~Kisner\inst{8}\orcid{0000-0003-3510-7134}\and
A.~Kremin\inst{8}\orcid{0000-0001-6356-7424}\and
M.~Landriau\inst{8}\orcid{0000-0003-1838-8528}\and
L.~Le~Guillou\inst{28}\orcid{0000-0001-7178-8868}\and
M.~Manera\inst{29,30}\orcid{0000-0003-4962-8934}\and
R.~Miquel\inst{31,30}\and
J.~Moustakas\inst{32}\orcid{0000-0002-2733-4559}\and
S.~Nadathur\inst{17}\orcid{0000-0001-9070-3102}\and
W.~J.~Percival\inst{33,34,35}\orcid{0000-0002-0644-5727}\and
I.~P\'erez-R\`afols\inst{36}\orcid{0000-0001-6979-0125}\and
F.~Prada\inst{37}\orcid{0000-0001-7145-8674}\and
G.~Rossi\inst{38}\and
E.~S\'anchez\inst{39}\orcid{0000-0002-9646-8198}\and
D.~Schlegel\inst{8}\and
R.~Sharples\inst{40,41}\orcid{0000-0003-3449-8583}\and
J.~H.~Silber\inst{8}\orcid{0000-0002-3461-0320}\and
D.~Sprayberry\inst{7}\and
G.~Tarl\'e\inst{26}\orcid{0000-0003-1704-0781}\and
H.~Zou\inst{42}\orcid{0000-0002-6684-3997}
}

\institute{
Institute of Space Sciences (ICE-CSIC), Campus UAB, Carrer de Can Magrans, s/n, E-08193 Barcelona, Spain
\and Department of Physics, Kathmandu University, Dhulikhel 45200, Nepal
\and Department of Physics \& Astronomy, University of Kansas,1251 Wescoe Hall Drive, Lawrence, KS 66045, USA
\and Institut d'Estudis Espacials de Catalunya (IEEC), 08860 Castelldefels (Barcelona), Spain
\and Department of Physics \& Astronomy, University of Rochester, 500 Joseph C.~Wilson Blvd., Rochester, NY 14627, USA
\and McWilliams Center for Cosmology and Astrophysics, Department of Physics, Carnegie Mellon University, Pittsburgh, PA 15213, USA
\and NSF NOIRLab, 950 N. Cherry Ave., Tucson, AZ 85719, USA
\and Lawrence Berkeley National Laboratory, 1 Cyclotron Road, Berkeley, CA 94720, USA
\and Department of Physics, Boston University, 590 Commonwealth Avenue, Boston, MA 02215 USA
\and Dipartimento di Fisica ``Aldo Pontremoli'', Universit\`a degli Studi di Milano, Via Celoria 16, I-20133 Milano, Italy
\and INAF-Osservatorio Astronomico di Brera, Via Brera 28, 20122 Milano, Italy
\and Department of Physics \& Astronomy, University College London, Gower Street, London, WC1E 6BT, UK
\and Instituto de F\'{\i}sica, Universidad Nacional Aut\'{o}noma de M\'{e}xico,  Circuito de la Investigaci\'{o}n Cient\'{\i}fica, Ciudad Universitaria, Cd. de M\'{e}xico  C.~P.~04510,  M\'{e}xico
\and University of California, Berkeley, 110 Sproul Hall \#5800 Berkeley, CA 94720, USA
\and Departamento de F\'isica, Universidad de los Andes, Cra. 1 No. 18A-10, Edificio Ip, CP 111711, Bogot\'a, Colombia
\and Observatorio Astron\'omico, Universidad de los Andes, Cra. 1 No. 18A-10, Edificio H, CP 111711 Bogot\'a, Colombia
\and Institute of Cosmology and Gravitation, University of Portsmouth, Dennis Sciama Building, Portsmouth, PO1 3FX, UK
\and University of Virginia, Department of Astronomy, Charlottesville, VA 22904, USA
\and Fermi National Accelerator Laboratory, PO Box 500, Batavia, IL 60510, USA
\and The University of Texas at Austin, Department of Astronomy, USA
\and Center for Cosmology and AstroParticle Physics, The Ohio State University, 191 West Woodruff Avenue, Columbus, OH 43210, USA
\and Department of Physics, The Ohio State University, 191 West Woodruff Avenue, Columbus, OH 43210, USA
\and The Ohio State University, Columbus, 43210 OH, USA
\and School of Mathematics and Physics, University of Queensland, Brisbane, QLD 4072, Australia
\and Department of Physics, University of Michigan, 450 Church Street, Ann Arbor, MI 48109, USA
\and University of Michigan, 500 S. State Street, Ann Arbor, MI 48109, USA
\and Department of Physics, Southern Methodist University, 3215 Daniel Avenue, Dallas, TX 75275, USA
\and Sorbonne Universit\'{e}, CNRS/IN2P3, Laboratoire de Physique Nucl\'{e}aire et de Hautes Energies (LPNHE), FR-75005 Paris, France
\and Departament de F\'{i}sica, Serra H\'{u}nter, Universitat Aut\`{o}noma de Barcelona, 08193 Bellaterra (Barcelona), Spain
\and Institut de F\'{i}sica d’Altes Energies (IFAE), The Barcelona Institute of Science and Technology, Edifici Cn, Campus UAB, 08193, Bellaterra (Barcelona), Spain
\and Instituci\'{o} Catalana de Recerca i Estudis Avan\c{c}ats, Passeig de Llu\'{\i}s Companys, 23, 08010 Barcelona, Spain
\and Department of Physics and Astronomy, Siena University, 515 Loudon Road, Loudonville, NY 12211, USA
\and Department of Physics and Astronomy, University of Waterloo, 200 University Ave W, Waterloo, ON N2L 3G1, Canada
\and Perimeter Institute for Theoretical Physics, 31 Caroline St. North, Waterloo, ON N2L 2Y5, Canada
\and Waterloo Centre for Astrophysics, University of Waterloo, 200 University Ave W, Waterloo, ON N2L 3G1, Canada
\and Departament de F\'isica, EEBE, Universitat Polit\`ecnica de Catalunya, c/Eduard Maristany 10, 08930 Barcelona, Spain
\and Instituto de Astrof\'{i}sica de Andaluc\'{i}a (CSIC), Glorieta de la Astronom\'{i}a, s/n, E-18008 Granada, Spain
\and Department of Physics and Astronomy, Sejong University, 209 Neungdong-ro, Gwangjin-gu, Seoul 05006, Republic of Korea
\and CIEMAT, Avenida Complutense 40, E-28040 Madrid, Spain
\and Centre for Advanced Instrumentation, Department of Physics, Durham University, South Road, Durham DH1 3LE, UK
\and Institute for Computational Cosmology, Department of Physics, Durham University, South Road, Durham DH1 3LE, UK
\and National Astronomical Observatories, Chinese Academy of Sciences, A20 Datun Road, Chaoyang District, Beijing, 100101, P.~R.~China
}

\date{Received \today; accepted ---}

\abstract{
The Tully--Fisher (TF) relation is an empirical tool for estimating distances to spiral galaxies. The Dark Energy Spectroscopic Instrument (DESI) Peculiar Velocity (PV) Survey uses \texttt{`tractor'} photometric position angles (PAs) to place its fibers along the semi-major axis of the galaxy and infer maximum rotational velocities for $\approx$ 53,000 spiral galaxies. Systematic errors may arise if the photometric PA differs from the kinematic PA measured from velocity fields.}
{We quantify the systematic uncertainty in DESI-TF distance estimates introduced by the use of photometric PAs, and assess the impact of photometric--kinematic PA misalignment on derived distances.}
{We analyze a curated sample of 215 nearby galaxies from the PISCO and AMUSING surveys. For consistency with the DESI-PV survey, we estimate the maximum rotational velocity at $0.4\,R_{26}$. We estimate TF distance using three different sets of parameters. We obtain kinematic parameters using \texttt{`PaFit'} from \texttt{`Cappellari Software'}. For global photometric parameters, we rely on data available on the Siena Galaxy Atlas SGA-2020 from \texttt{`tractor'} photometry, and we also use \texttt{`HostPhot'} photometry as an additional photometric approach. } 
{Approximately $28\%$ of our sample exhibit photometric--kinematic PA misalignment $>10^\circ$. The $\mathrm{median}$ bias in distance is $\mathrm{\approx +1.98\,Mpc}$ with a skewed residual distribution dominated by outliers. The overall standard deviation ($\sigma$) for the estimated distance is $19.35\,\mathrm{Mpc}$. The misaligned galaxies show twice the dispersion compared to aligned galaxies. The Mean Percentage Error (MPE) is $2.8 \% \pm 1.96$ (SE) throughout the sample. }
{Kinematic constraints from IFS reveal a low positive median bias $\approx1.98 \,\mathrm{Mpc}$ but high galaxy-to-galaxy scatter in DESI-TF distances particularly for misaligned galaxies. Because DESI-TF targets lack kinematic PA measurements, we recommend a global fractional uncertainty of $\approx 3.5\%$ for DESI-TF distances. When kinematic information is available, aligned galaxies with 
photometric--kinematic PA offsets $<10^\circ$ are consistent with $\approx 1\%$ uncertainty, while strongly misaligned or kinematically incomplete systems warrant an upper threshold of $\approx 10\%$. 
}
\maketitle

\section{Introduction}
Understanding the structure and evolution of nearby spiral galaxies is a central pursuit in extragalactic astronomy. It forms the foundational studies of star formation, galactic dynamics, and the distribution of dark matter \citep{ 1986RSPTA.320..447V, Sofue_2001}. A key requirement for these studies is an accurate determination of extragalactic distances, which enables the calibration of galaxy properties and the mapping of large-scale cosmic flows. For spiral galaxies, the Tully–Fisher (TF) relation serves as a widely adopted method for estimating distances.

The conceptual foundation of the TF relation dates back to the early 20th century. \citet{1922ApJ....55..406O} first demonstrated that the rotational velocity and absolute magnitude of the Andromeda galaxy can be used to infer its distance. This approach was subsequently refined and generalized, culminating in the formalization of the TF relation in 1977 \citep{1977A&A....54..661T}. The TF relation established that a galaxy’s luminosity scales as a power law of its rotational velocity, $L \propto V^{\alpha}$, with $\alpha$ typically between 3 and 4. Over the decades, the TF relation has been extensively used and improved, with methodological advances in both measurement of rotational velocities (from HI line widths to modern spectroscopic rotation curves) and the calibration itself \citep{1992ApJ...387...47P, Giovanelli_1998, Tully_2000,  Kourkchi_2020, Khaled_said_2023}. Reliable application of the TF relation requires precise sample calibration and rigorous control of systematic uncertainties \citep{Willick_1997,2012_Tully_calibration,socre_2013,2012courtois_calibration, Kourkchi_2020, Khaled_said_2023, douglass2025desiedrcalibratingtullyfisher}.

Leveraging these foundations, recent studies have introduced iterative calibration schemes that use cluster samples together with independent distance anchors to minimize error propagation \citep{Kourkchi_2020}. These schemes repeatedly refine the TF zero-point while anchoring it to external distance indicators, thereby suppressing the accumulation of systematic offsets along the distance ladder. With this stabilized calibration framework, the Dark Energy Spectroscopic Instrument (DESI) is carrying out the most competitive TF measurement to date, targeting $\sim$53{,}000 spiral galaxies to constrain Peculiar Velocity (PV) \citep{2023MNRAS.525.1106S, douglass2025desiedrcalibratingtullyfisher}. 

DESI is a robotic, fiber-fed, highly multiplexed spectroscopic surveyor that operates on the Mayall 4-meter telescope at Kitt Peak National Observatory \citep{DESI2022.KP1.Instr}. It can obtain simultaneous spectra of almost 5000 objects over a $\sim3^o$ field \citep{ Corrector.Miller.2023, FiberSystem.Poppett.2024}. It is conducting an eight-year survey of about $17,000 \mathrm{deg}^2$ of the sky. The full survey will lead to 63 million spectroscopically-confirmed galaxies and quasars, compared to initial forecasts of 39 million \citep{DESI2016b.Instr}. The sheer scale of the DESI experiment necessitates multiple supporting software pipelines and products \citep{Spectro.Pipeline.Guy.2023,SurveyOps.Schlafly.2023}. Cosmological constraints from DESI’s First Data Release (DR1) were derived from the KP7B full‑shape analysis \citep[DR1][]{DESI2024.I.DR1}. Work toward cosmological results from the Second Data Release (DR2) is now underway {DESI.DR2.DR2}.

The DESI-PV Survey is a secondary DESI program targeting galaxies at $z < 0.15$ to obtain redshift–independent distances across $14{,}000\,\mathrm{deg}^2$. It measures distances for $\sim133{,}000$ early-type galaxies using the Fundamental Plane (FP) relation and for $\sim53{,}000$ late-type galaxies using the TF relation. When combined with DESI’s precise spectroscopic redshifts, these distance estimates yield one of the largest homogeneous PV samples to date, enabling improved constraints on the local matter distribution and the growth of structure \citep{2023MNRAS.525.1106S}.

For the TF sample, DESI's spectroscopic fibers are positioned along the photometric major axis as determined by the automated photometric pipeline \texttt{`tractor'} on Legacy Survey images \citep{Moustakas_2023}. The rotational velocities are derived from spectra at the fiber positions. The photometric PA (hereafter, ${\rm PA}_{\rm phot}$) usually aligns within 5-10 \% to the kinematic PA (hereafter, ${\rm PA}_{\rm kin}$) \citep{Barnes_2003}. However, potential misalignment between ${\rm PA}_{\rm phot}$ and ${\rm PA}_{\rm kin}$ can occur due to bars and non axisymmetric structures, galaxy interaction and mergers, triaxial stellar distribution, gas accretion or inflows projection effect and dynamical effects \citep{Toomre_1972, Veilux_2002, Rupke_2010, Barrera_Ballesteros_2014, Bloom_2017,  Yang_2019}. Agreement between these PAs is conceptually important for TF work because the inclination correction and the measured velocity are assumed to refer to the same projected disk orientation. When the two PAs differ, the resulting mismatch leads to different deprojections of the rotation field and a systematic offset in the inferred TF velocity. 

Although DESI does not perform full tilted--ring modeling, the same geometric projection used in the tilted--ring formalism \citep{1989A&A...223...47B}  applies to any line-of-sight velocity measured at the sky position sampled by a fiber. In this geometry, the observed velocity is related to the intrinsic rotational velocity by
\begin{equation}\label{eq:projection}
V_{\rm obs} = V_{\rm rot}\,\sin i\,\cos\theta,
\end{equation}
where $\theta$ is the angular offset between the fiber PA and the true kinematic major axis.  As the fiber is aligned using ${\rm PA}_{\rm phot}$ instead of ${\rm PA}_{\rm kin}$, then $\theta$ becomes $\Delta{\rm PA}$, with $\Delta{\rm PA} = {\rm PA}_{\rm phot}-{\rm PA}_{\rm kin}$. The equation becomes:
\begin{equation}
V_{\rm obs} = V_{\rm rot}\,\sin i\,\cos(\Delta{\rm PA}),
\end{equation}
This shows that if $\Delta{\rm PA} > 0$, $\cos{\theta} \neq 1$,  which means PA misalignment reduces the observed rotational velocity. This introduces a key systematic uncertainty not only in the maximum rotational velocity but also in the inferred distance. This issue is particularly acute in automated surveys like DESI, where detailed kinematic measurements are unavailable for all targets. We quantify these systematic uncertainties in DESI-TF distances induced by photometric--kinematic misalignment and assess their impact on the derived distances.

In this work, we adopt a logarithmic TF formulation in Eq.~\ref{eq:TF_relation} to calibrate the relation and derive galaxy distances from rotational velocity measurements.
\begin{equation} \label{eq:TF_relation}
    M = a \, (\log V_{\text{rot}} - 2.5) + b,
\end{equation}
where $V_{\text{rot}}$ is the inclination-corrected rotational velocity evaluated at $0.4\,R_{26}$. Here, $0.4\,R_{26}$ denotes the $r$-band semi-major axis corresponding to the 26~mag~arcsec$^{-2}$ isophote. The parameters $a$ and $b$ are the slope and zero point of the TF relation, respectively \citep{sakai_2000,Tully_2000,Kourkchi_2020,2023MNRAS.525.1106S,2024AAS...24340709D, douglass2025desidr1peculiarvelocity}.

Photometric--kinematic misalignment and its potential impact on TF distances have been discussed qualitatively in earlier work \citep[e.g.][]{Barnes_2003,Barrera_Ballesteros_2014, Bloom_2017}. Integral-field spectroscopy (IFS) has previously been used to study TF relations and velocity fields, but mainly to optimise kinematic measurements rather than to emulate survey-specific observational constraints. The novelty of our work lies in explicitly framing the PA-misalignment problem within the context of the DESI-PV survey, which employs \rm PA$_{\rm phot}$ and a three-fiber configuration (one central fiber and two fibers at $0.4\,R_{26}$ on either end) aligned to the photometric major axis. We use a curated IFS sample of 215 PMAS-PPak Integral-field Supernova hosts COmpilation PISCO; \citep{2018ApJ...855..107G} and All-weather MUSE Integral-field Nearby Galaxies AMUSING; \cite{2016MNRAS.455.4087G} survey to quantify how PA offsets between photometric and kinematic axes propagate into TF distances. This approach directly connects the IFS-based PA-alignment literature with the broader TF systematics studies \citep{Willick_1997, socre_2013, Kourkchi_2020, Khaled_said_2023}. They focused on photometric corrections, inclination uncertainties, and internal extinction, but have not incorporated kinematic PA information. Our analysis provides the first empirical estimate of PA-induced velocity and distance systematics tailored to the DESI-PV survey, establishing the magnitude of PA-related biases relevant for DESI TF distance estimates.

This paper is organized as follows: Section \ref{data} describes the data sources and sample selection; Section \ref{methods} outlines the methods for photometric and kinematic analysis; Section \ref{results} presents results, misalignment statistics, and their impact on TF distances; and Section \ref{conclusion} discusses implications for DESI and future surveys.

\section{Observations and data}\label{data}

We use a curated sample of 215 nearby galaxies with integral field spectroscopy (IFS) drawn from the PISCO \citep{2018ApJ...855..107G} and AMUSING \citep{2016MNRAS.455.4087G} surveys. We complement these kinematic measurements with homogeneous photometric parameters from the Siena Galaxy Atlas (SGA-2020; \citealt{Moustakas_2023}), which itself is based on deep optical imaging from the Legacy Surveys (Data Release~9; \citealt{Dey_2019}). The combination of high-quality velocity maps and consistent photometry provides a robust basis for TF analysis and DESI target selection. Additionally, we run the \texttt{`HostPhot'}\footnote{\href{https://hostphot.readthedocs.io/}{https://hostphot.readthedocs.io/}} pipeline \citep{Müller-Bravo2022} on Panoramic Survey Telescope and Rapid Response System (Pan‑STARRS1, hereafter PS1; \citealt{chambers2019panstarrs1}) $r$-band cutouts centered on each galaxy to obtain ancillary photometric parameters to have independent photometric checks and a comparison set with \texttt{`tractor'}. PS1 is the first telescope of the Pan‑STARRS survey system. It is a wide‑field 1.8‑m optical telescope equipped with a 1.4‑gigapixel camera that conducts imaging in five broadband filters ($g,r, i,z,y$).

\subsection{The Legacy Surveys}\label{sec:legacy_survey}

The Legacy Surveys, formerly known as the DESI Legacy Imaging Surveys, were originally designed to image approximately $14,000\, \mathrm{deg}^2$ of extragalactic sky in three optical bands ($g$, $r$, $z$) to support target selection for the DESI. Subsequent observing and reprocessing expanded the available imaging to nearly $20,000\, \mathrm{deg}^2$ in later data releases (e.g., DR8). To achieve wide-area coverage within a limited time frame, the imaging is conducted using three complementary telescope platforms: the Blanco 4m Telescope at Cerro Tololo Inter-American Observatory (CTIO) \citep{2015AJ....150..150F}, the Mayall 4m Telescope at Kitt Peak National Observatory (KPNO), \citep{Dey_2019}, and the Bok 2.3m Telescope at KPNO, \citep{zou}. Together, these surveys form a unified imaging footprint with sufficient depth, resolution, and photometric uniformity to enable robust DESI target selection—particularly in regions where previous surveys such as Sloan Digital Sky Survey (SDSS) and Pan-STARRS1 were too shallow or incomplete.

In addition to optical imaging, the Legacy Surveys incorporate mid-infrared photometry from the Wide-field Infrared Survey Explorer (WISE; \citealt{Wright_2010A}) and Near-Earth Object Wide-field Infrared Survey Explorer (NEOWISE; \citealt{Mainzer_2014}) reactivation missions. These data are coadded into deep image stacks and matched to optical sources using the \texttt{`tractor'} modeling code, which performs multi-band photometric fitting across variable point-spread functions and CCD geometries \citep{2016ascl.soft04008L}.

\subsection{Siena Galaxy Atlas}

Building on the deep, wide-area optical imaging provided by the Legacy Surveys described in Sect.~\ref{sec:legacy_survey}, the Siena Galaxy Atlas (SGA-2020) is a curated catalog of large angular-diameter galaxies constructed through catalog compilation, photometric modeling, and completeness validation, comprising 383,620 galaxies \citep{Moustakas_2023}. SGA-2020 is estimated to be more than 95\% complete for galaxies with angular diameters $R(26) \geq 25\arcsec $ and $r < 18$, based on comparison with the Homogenized Extended Catalog of Galaxies (HECATE) \citep{Kovlakas_2021}. The catalog provides precise coordinates, multi-wavelength image mosaics, azimuthally averaged surface brightness and color profiles, \texttt{`tractor'} model images, photometry, and extensive ancillary metadata \citep{Moustakas_2023}. Roughly $70\%$ of the SGA galaxies lie within the DESI footprint, and the DESI-PV Survey aims to derive TF distance measurements for about 53,000 galaxies in the SGA sample \citep{2023MNRAS.525.1106S}. 

\subsection{IFS observations and velocity-field construction} \label{sec:halpha_vfield}

Integral Field Spectroscopy (IFS) provides spatially resolved spectroscopic information for galaxies. Although Mapping Nearby Galaxies at APO (MaNGA) \citep{Bundy_2015} includes IFU data for $\sim$10,000 nearby galaxies, we instead use data from the PISCO \citep{2018ApJ...855..107G} and AMUSING \citep{2016MNRAS.455.4087G} surveys. This choice is motivated by the significantly higher spatial resolution of AMUSING (MUSE seeing $\sim0.6$--$1.0\arcsec$), and the larger spatial extent covered by PISCO, both of which are essential for resolving inner velocity fields and obtaining robust kinematic position angle measurements. While PISCO does not surpass MaNGA in spatial resolution, being limited by $2.7\arcsec$ fibers, it provides extensive coverage of nearby SN host galaxies. Together, PISCO and AMUSING form targeted IFU samples that enable a consistent zero-point calibration of the TF relation while minimizing beam-smearing effects in the kinematic analysis.

To construct the velocity maps, the stellar continuum is modeled and subtracted using spectral population synthesis techniques (e.g., \textsc{STARLIGHT} \citep{cid_starlight_2005}, or the Pipe3D framework \citep{2016A&A...594A..36S}), removing underlying Balmer absorption features and isolating the nebular emission. Emission lines are fitted in the observed frame, and velocities are obtained from the shift of the H$\alpha$ centroid relative to its rest wavelength, producing two-dimensional maps of ionized-gas kinematics. \citep{2016MNRAS.455.4087G, 2018ApJ...855..107G, Lopez_2020}. These maps serve as the input for our \texttt{`kinemetry'} analysis, from which we derive kinematic parameters and measure the maximum rotational velocity along axes defined by three independent PAs: (\texttt{`HostPhot'}, \texttt{`kinemetry'}, and \texttt{`tractor'}). 

\subsection{Target selection}\label{sec:sam}

Our sample is constructed through a multi-stage process aimed at ensuring high-quality kinematic and photometric measurements. The initial dataset comprised 986 galaxies with integral field spectroscopy (IFS) observations drawn from the PISCO (610 galaxies) and AMUSING (376 galaxies) surveys. Due to the southern sky concentration of the AMUSING sample, we applied a declination cut at $\delta > -20^\circ$, retaining only 50 galaxies from AMUSING. We then conducted a visual inspection of the velocity maps, selecting galaxies that exhibited coherent rotation patterns and sufficient spatial coverage for reliable kinematic modeling. This step yields a total of 354 galaxies—320 from PISCO and 34 from AMUSING.

Subsequently, we cross-match these galaxies with the Siena Galaxy Atlas (SGA-2020) to ensure consistent photometric data. Sixteen galaxies lacked SGA counterparts and are excluded, reducing the working sample to 338 galaxies. The PISCO/AMUSING galaxies missing from SGA‑2020 are primarily small or compact systems that fall below the SGA angular‑diameter threshold, or objects with incomplete or unreliable HyperLeda metadata. A smaller number are excluded due to imaging artifacts, shredding, or incomplete DR9 coverage. A series of quality control criteria is further applied to ensure robustness in both photometric and kinematic measurements. We exclude 51 galaxies for which PAs could not be reliably measured by at least one of our three methods (see Sect.~\ref{sec:phot_analysis}). An additional 8 galaxies are discarded due to co-addition failures during \textsc{HostPhot} processing, often caused by foreground stars or imaging artifacts. We also remove 55 galaxies with NaN values in their velocity fields at $0.4\,R_{26}$ due to poor data quality, and 4 galaxies with significant interpolation errors during the \texttt{`kinemetry'} fitting process. Finally, five galaxies with redshift $z > 0.04$ are excluded to ensure consistency with local TF calibrations. Although DESI, PISCO, and AMUSING are capable of measuring galaxies beyond $z = 0.04$, only a few objects in our sample lie at higher redshift, so excluding those systems does not materially affect our calibration.

\begin{figure}[t]
    \centering
    \includegraphics[width=\linewidth]{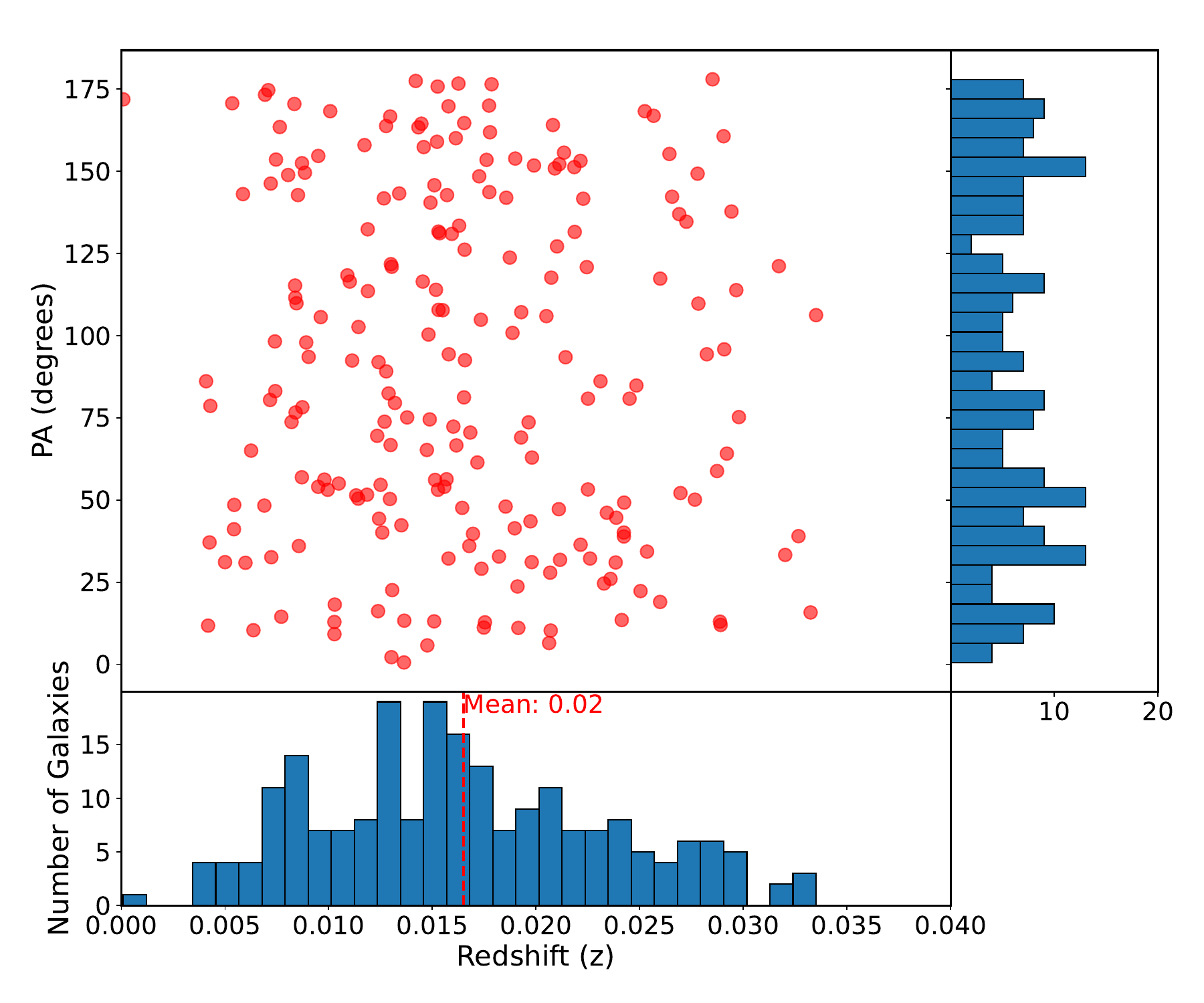}
    \caption{The distribution of PA as a function of redshift for the target sample. A redshift cut of $z<0.04$ is applied, resulting in a mean redshift of 0.02. The PA distribution indicates that no specific range of PA values is restricted, with the sample covering the full range from 0 to 180 degrees.}
    \label{fig:redshift_PA}
\end{figure}

\begin{figure}[t]
    \centering
    \includegraphics[width=\linewidth]{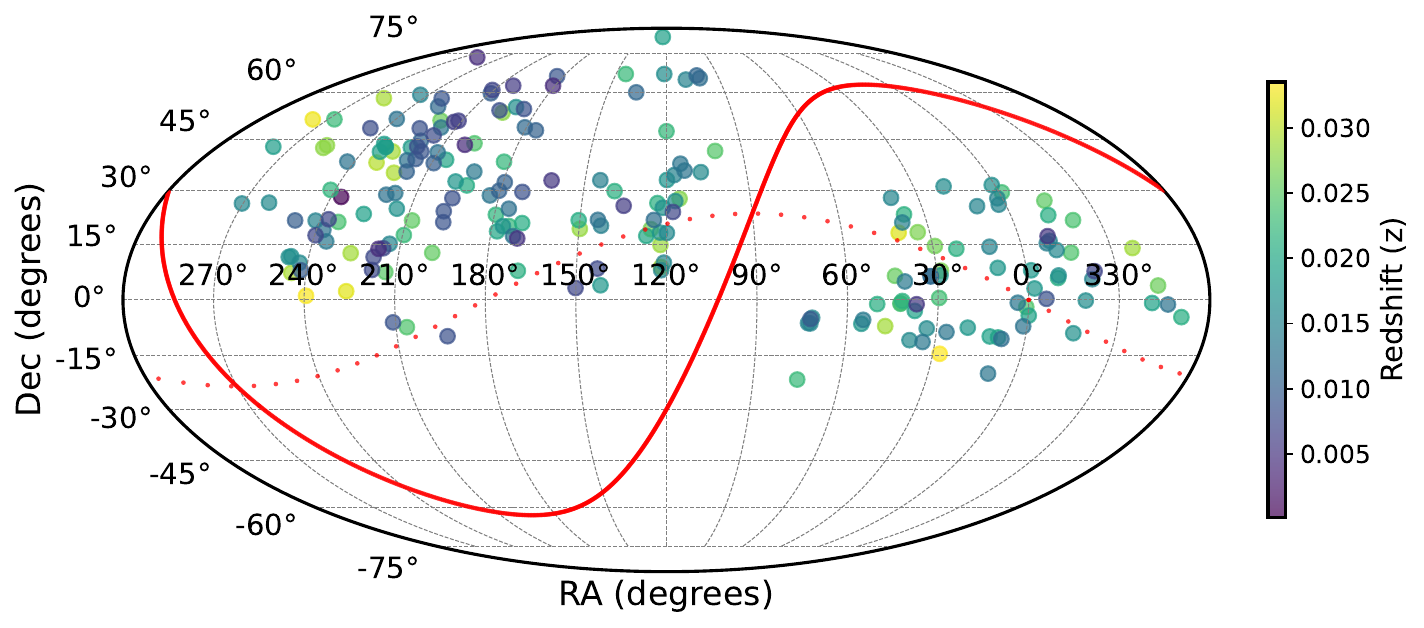}
    \caption{Sky distribution of galaxies in equatorial coordinates, color-coded by redshift. The Mollweide projection centers on RA = 120\textdegree, with each point representing a galaxy from the sample. The red dotted curve marks the RA = 0\textdegree\ meridian as defined by DESI's plotting convention. This footprint lies within the spatial footprint of the DESI-PV survey.}
    \label{fig:Ra_Dec_z_2D}
\end{figure}

Following all selection steps, the final sample consists of 215 galaxies: 203 from PISCO and 12 from AMUSING. The sample is restricted to redshifts $z < 0.04$, with a median redshift of $z \approx 0.02$. As illustrated in Fig.~\ref{fig:redshift_PA}, the sample spans the full range of PAs between $0^\circ$ and $180^\circ$, with no imposed selection bias in PA distribution. To validate the angular coverage of our sample, we compare its sky distribution to the DESI DR1 mock BGS catalogs constructed using the nested-HOD model described by \cite{Bautista_2025}. As shown in Fig.~\ref{fig:Ra_Dec_z_2D}, our sample exhibits a spatial footprint that closely matches the DESI DR1 coverage, both in equatorial coordinates and within the redshift range covered by DESI. The DESI TF program is designed to target late-type spiral galaxies suitable for TF analysis, but the full DESI DR1 TF catalog contains a broader morphological mix (68.9\% spirals, 24.5\% irregulars, and small fractions of lenticular, elliptical, and unclassified systems), reflecting the diversity of the full survey \citep{DESI2024.I.DR1}. Our SN-host IFS sample overlaps substantially with the dominant spiral population, though it should not be regarded as fully representative of the complete DESI TF sample; SN hosts, particularly CCSN hosts, may exhibit enhanced star formation and interaction rates. Quantitatively, our IFS sample is more spiral-dominated (93\% spirals) than the DESI DR1 TF catalog (68.9\% spirals) and is restricted to a lower redshift ($z < 0.04$ compared to $z \lesssim 0.15$ for DESI). Although the sky footprint and dominant spiral population are well matched, our sample does not capture the full morphological diversity or higher-redshift tail of the DESI TF survey. Consequently, the PA misalignment fraction measured here should be regarded as an upper limit on the systematics present in the broader DESI TF sample.

\section{Methods}\label{methods}

We quantify systematic uncertainties in TF distances by comparing three independent approaches to measuring galactic rotation properties: 
(1) DESI's baseline SGA-\texttt{`tractor'} photometric method \citep{Moustakas_2023}, 
(2) An alternative \texttt{`HostPhot'} photometric method \citep{Müller-Bravo2022} using PS1 imaging; and 
(3) A kinematic reference method using \texttt{`kinemetry'} \citep{2006MNRAS.366..787K} on H$\alpha$ velocity fields. 

All methods utilize common SGA-2020 parameters (isophotal radii, $R_{26}$, and $r$-band magnitudes within those radii, $m_r(R_{26})$) but independently determined PAs, inclination angles $(i)$, and axis ratios $(b/a)$. For each method, we identify saturation velocities, measure rotation velocities along each galaxy's defined major axis at $0.4\,R_{26}$, and correct for inclination \citep{1958MeLuS.136....1H} using geometric projection in Eq~\ref{eq:projection}. The inclination is derived from the photometric axis ratio via:
\begin{equation}
\cos^2 i = \frac{(b/a)^2 - q_0^2}{1 - q_0^2},
\end{equation}
where $q_0 = 0.2$ is the intrinsic disk thickness \citep{Tully_2000}.

Our choice of extracting velocities at $0.4\,R_{26}$ follows the DESI-PV observing strategy. During Survey Validation (SV), DESI placed fibers at $0.33\,R_{26}$, $0.67\,R_{26}$, and $R_{26}$ along the photometric major axis to determine how far from the galaxy center reliable redshift measurements could be obtained. The SV results showed that redshift success rates remained high at $0.33\,R_{26}$ but dropped sharply beyond this radius. The success rate declines to 28.3\% at 0.66\,$R_{26}$ and only 12.8\% at $R_{26}$ \citep{2023MNRAS.525.1106S}. They also considered the fact that the velocities measured at $0.33\,R_{26}$ remain too close to the rising part of the rotation curve for TF work. Combining these considerations, DESI adopted $ 0.4\,R_{26}$ as the fiducial radius for the main survey \citep{2023MNRAS.525.1106S, douglass2025desidr1peculiarvelocity}. We therefore extract IFS velocities at $0.4\,R_{26}$. This choice also ensures that the velocities are measured at a consistent radius across all galaxies, maintaining methodological uniformity. 

\subsection{Photometric analysis}\label{sec:phot_analysis}

We use photometric PAs from two independent photometric pipelines. First, we adopt the PA, axis ratio $(b/a)$, isophotal radius $R_{26}$, and apparent magnitude $m_r(26)$ from the \textsc{SGA-2020} catalogue, which is based on imaging from the \textsc{DESI} Legacy Surveys \citep{Moustakas_2023}. While \texttt{`tractor'} \citep{2016ascl.soft04008L} models each source within a tile using a unified likelihood optimisation, the \textsc{SGA-2020} pipeline includes additional steps like custom mosaic masking, residual-based masking, modified detection thresholds, and non-parametric surface brightness to improve photometric accuracy for bright, resolved galaxies \citep{Moustakas_2023}.

As an independent check, we run the \texttt{`HostPhot'} pipeline \citep{Müller-Bravo2022} on PS1 $r$-band cutouts centered on each galaxy. In HostPhot, we first identify and mask all foreground stars (using a PSF‐based detection threshold and cross‐matching with the Gaia catalogue) so that the subsequent elliptical isophote fitting only sees cleaned galaxy light. \texttt{`HostPhot'} then fits elliptical isophotes to determine the major‐axis PA and semi‐major axis length. Beyond the masking strategy, HostPhot uses a non-parametric isophotal fit in a single band, whereas \texttt{`tractor'} employs a parametric forward model that simultaneously fits all sources using Sérsic or exponential profiles. Because of these methodological differences, \texttt{`HostPhot'} sometimes yields a slightly different PA, particularly for galaxies with a bright star near the disk, than {\sc Tractor’s} forward-modeling approach. By comparing the PAs from \texttt{`HostPhot'} and \texttt{`tractor'}, we obtain two independent orientation estimates: one from a masked isophote fit \texttt{`HostPhot'} and another from a simultaneous multi‐source model \texttt{`tractor'}. A summary of the photometric differences between the \texttt{`HostPhot'} and \texttt{`tractor'} pipelines (PA, inclination, and $R_{26}$) is provided in Appendix \ref{app_tab: Photometric}, which illustrates the typical offsets between the two PAs.

\begin{figure}[!t]
\centering
\includegraphics[width=\columnwidth]{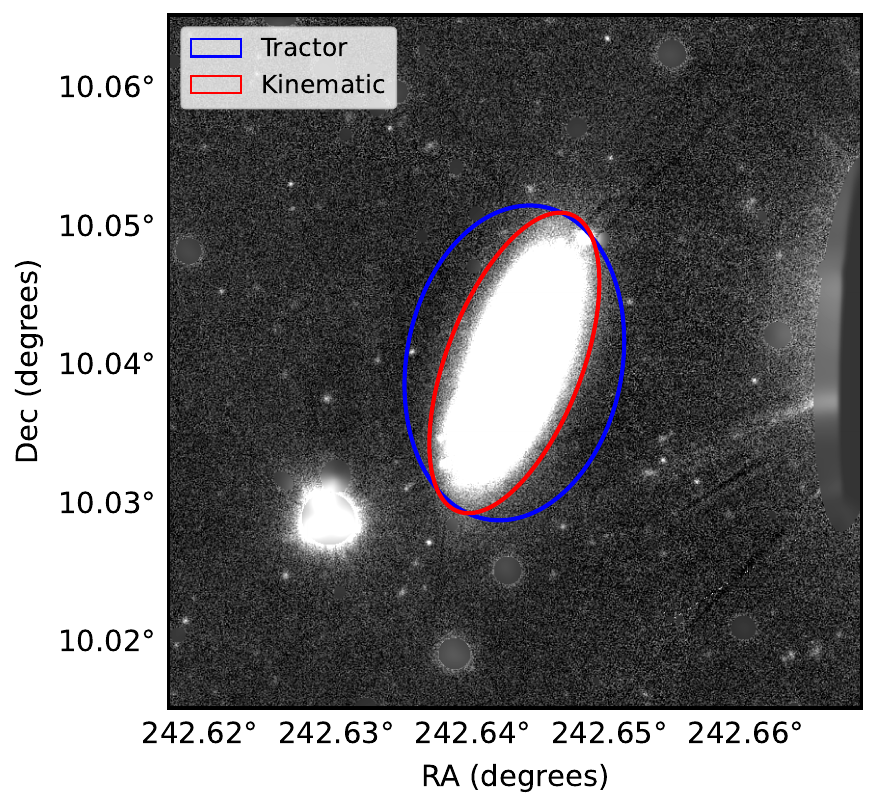}
\caption{Image cutout of IC~1199 from PS1, processed with \texttt{`HostPhot'}. Several foreground stars have been masked. We overlay two ellipses constructed using the \texttt{`HostPhot'} and \texttt{`tractor'} PA and inclination angle (i) to compare how each pipeline models the galaxy’s projected light distribution. The resulting $\Delta \mathrm{PA}$ of 11.55$^\circ$ makes this an outlier in our consistency check.}
\label{fig:hostphot}
\end{figure}

Fig.~\ref{fig:hostphot}  shows an illustrative example of the elliptical aperture fit produced by \texttt{`HostPhot'} and \texttt{`tractor'} for the outlier galaxy IC~1199. The PA produced by two photometric pipelines for this galaxy disagrees at the level of $\Delta{\rm PA} = 11.55^\circ$, just above our outlier threshold of $\Delta{\rm PA} \geq 10^\circ$. This same galaxy is used to illustrate Figs.~\ref {fig:hostphot}--\ref{fig:saturate}.  

\subsection{Kinematic analysis} \label{velifs} 

\begin{figure}[t]
\centering
\includegraphics[width=\columnwidth]{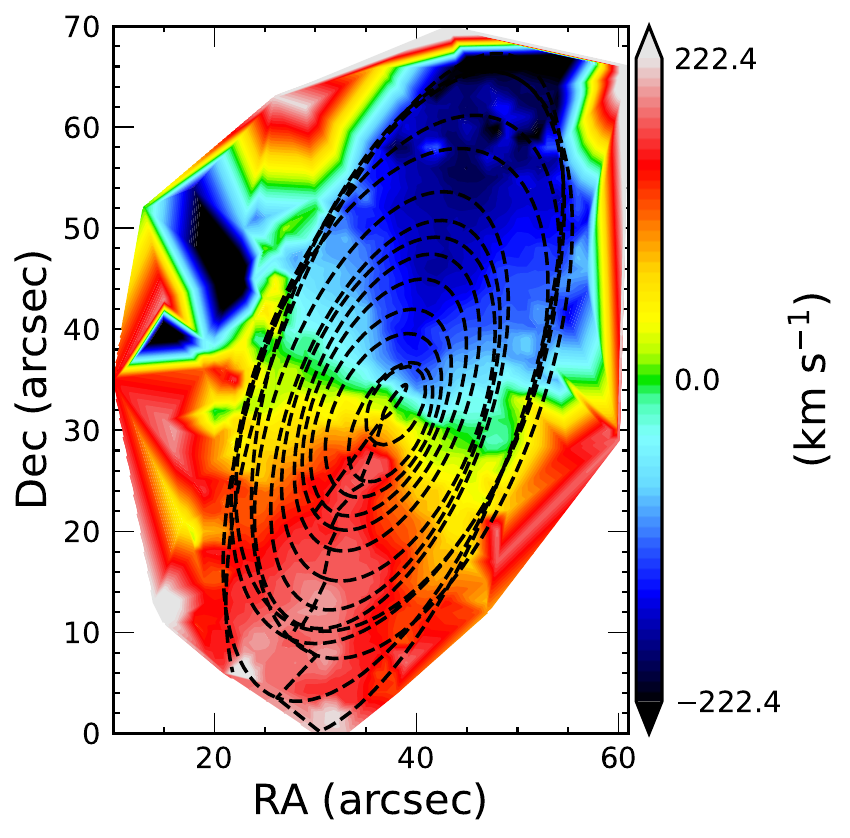}
\caption{Velocity field map overlaid with elliptical velocity rings. The color map represents the line-of-sight velocity in $\mathrm{km\ s^{-1}}$, where red indicates receding motion (positive velocities), and blue represents approaching motion (negative velocities) relative to the systemic velocity. The black dashed ellipses correspond to the fitted kinematic rings, representing different radii from the galaxy center. These ellipses are aligned with the galaxy's rotation axis, with their orientations adjusted by the PA, and their eccentricities scaled by the axis ratio $q$, reflecting the projected shape of the rotating disk. The velocity contours highlight the kinematic structure of the disk, with color variations illustrating velocity gradients.}
\label{fig:kin_vel_maps}
\end{figure}

\begin{figure}[t]
\centering
\includegraphics[width=\columnwidth]{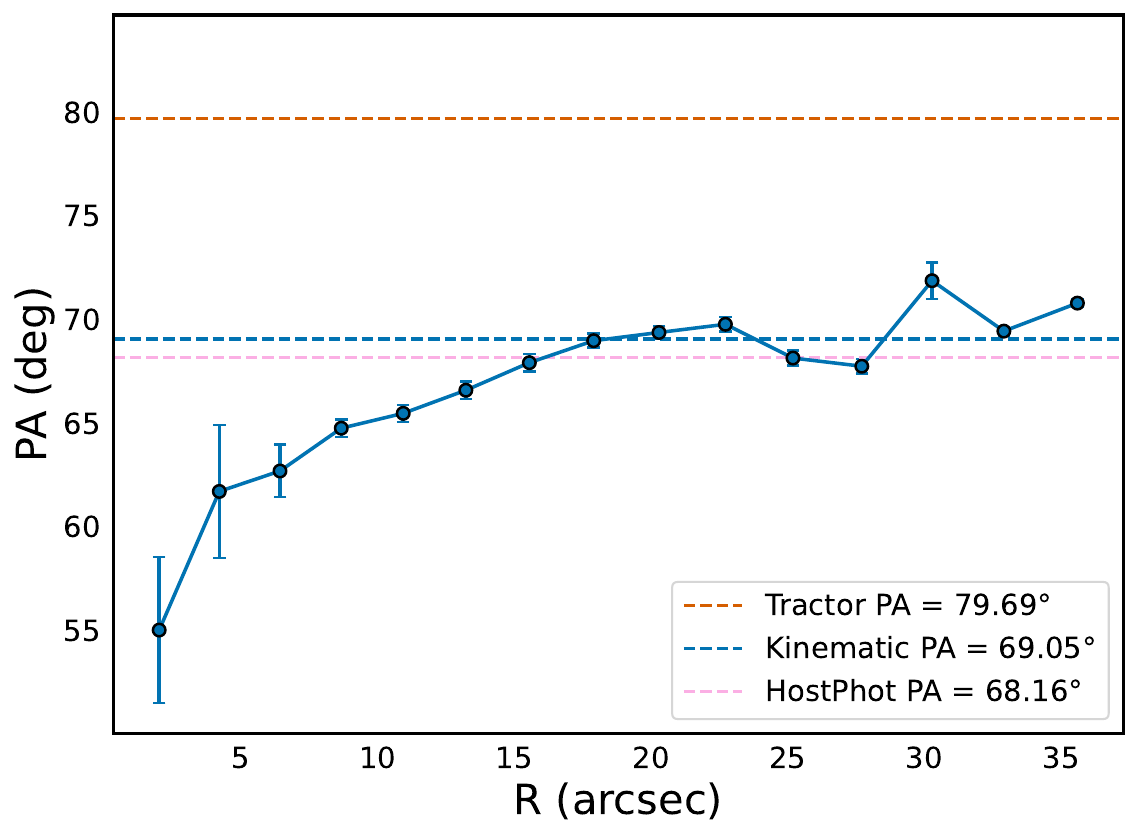}
\caption{PA profile derived from \texttt{`kinemetry'} analysis of IC\,1199. The PA (measured west of north) is shown as a function of radius R (arcsec) from the center. Each point corresponds to the semi-major axis orientation of an elliptical ring fit to the velocity map. The orange horizontal line represents the PA obtained from \texttt{`tractor'} photometric software, the pink horizontal line represents the \texttt{`HostPhot'} PA fit, and the blue fit represents PA$_{kin}$ from \texttt{`Pafit'} of Cappellari software. Here, the \texttt{`HostPhot'}s measurement is closer to the kinematic value, while the \texttt{`tractor'}s measurement deviates higher. } 
\label{fig:kin_pa_fit}
\end{figure}

Following previous IFS-based studies \citep{2014A&A...572A..38G,2016A&A...591A..48G,2016MNRAS.455.4087G}, we use the H$\alpha$ two-dimensional velocity fields from PISCO and AMUSING as an input for our kinematic analysis. Before running \texttt{`kinemetry'}, we visually inspect representative spaxel fits and check the FITS header metadata to verify the correct identification of the H$\alpha$ emission line. We analyze the internal kinematics using the \texttt{`kinemetry'} software \citep{2006MNRAS.366..787K}, which applies a generalized tilted-ring formalism \citep{1989A&A...223...47B} to the velocity map. The method expands the velocity field along a set of concentric elliptical rings using a truncated Fourier series:
\begin{equation}
K (\alpha, \psi) = A_0 + \sum_{n=1}^{N} \left( A_n(\alpha) \sin(n \psi) + B_n(\alpha) \cos(n \psi) \right)
\end{equation}
where $\psi$ is the eccentric anomaly, $\alpha$ is the semi-major axis length of the elliptical ring, and $N$ (typically 5–6) is the maximum harmonic order. For each ring, the ellipse geometry (center, PA, and flattening $q$, where $q = b/a$) is determined by minimizing the harmonic coefficients most sensitive to errors in the ellipse parameters. For even kinematic moments (e.g., surface-brightness maps), the optimal sampling ellipse is found by minimizing the low-order terms $A_1$, $B_1$, $A_2$, and $B_2$ \citep{10.1093/mnras/182.4.797}. For the first kinematic moment (the velocity field), Kinemetry instead determines the ellipse geometry by minimizing the terms most sensitive to PA and flattening errors, namely $A_1$, $A_2$, $B_2$, $A_3$, and $B_3$, which leads to the $\chi^2$ expression in Eq.~\ref{eq:chisquared}. The term $B_1$ is not included because it contains the bulk rotation signal rather than geometric misalignment.
\begin{equation}\label{eq:chisquared}
\chi^2 = A_1^2 + A_2^2 + B_2^2 + A_3^2 + B_3^2,
\end{equation}
A second stage refines the solution using a non-linear least-squares algorithm, fixing the center obtained from the first pass. This implementation uses the MINPACK Levenberg–Marquardt method \citep{Moré:126569}, yielding the final estimates for the kinematic center, ellipticity, and PA. Fig.~\ref{fig:kin_vel_maps} shows an example of the velocity-field decomposition for IC\,1199.

For each elliptical ring, the kinematic PA is calculated from the first-order harmonic terms as the orientation of the kinematic major axis, corresponding to the direction of maximum velocity gradient: 
\begin{equation}
\phi_{\rm PA} = \frac{1}{2} \arctan \left( -\frac{B_2}{A_2} \right),
\end{equation}
With the appropriate quadrant correction applied. This represents the projected orientation of the galaxy's angular momentum vector on the sky. Fig.~\ref{fig:kin_pa_fit} shows the resulting radial PA profile, where deviations from constancy indicate kinematic twists or disk warping. Complete \texttt{`kinemetry'} analysis plots for IC\,1199 are provided in Appendix \ref{app_sec:kinemetry}.

\subsection{Maximum rotational velocity}\label{satpoint}

\begin{figure}[t]
\centering
\includegraphics[width=\columnwidth]{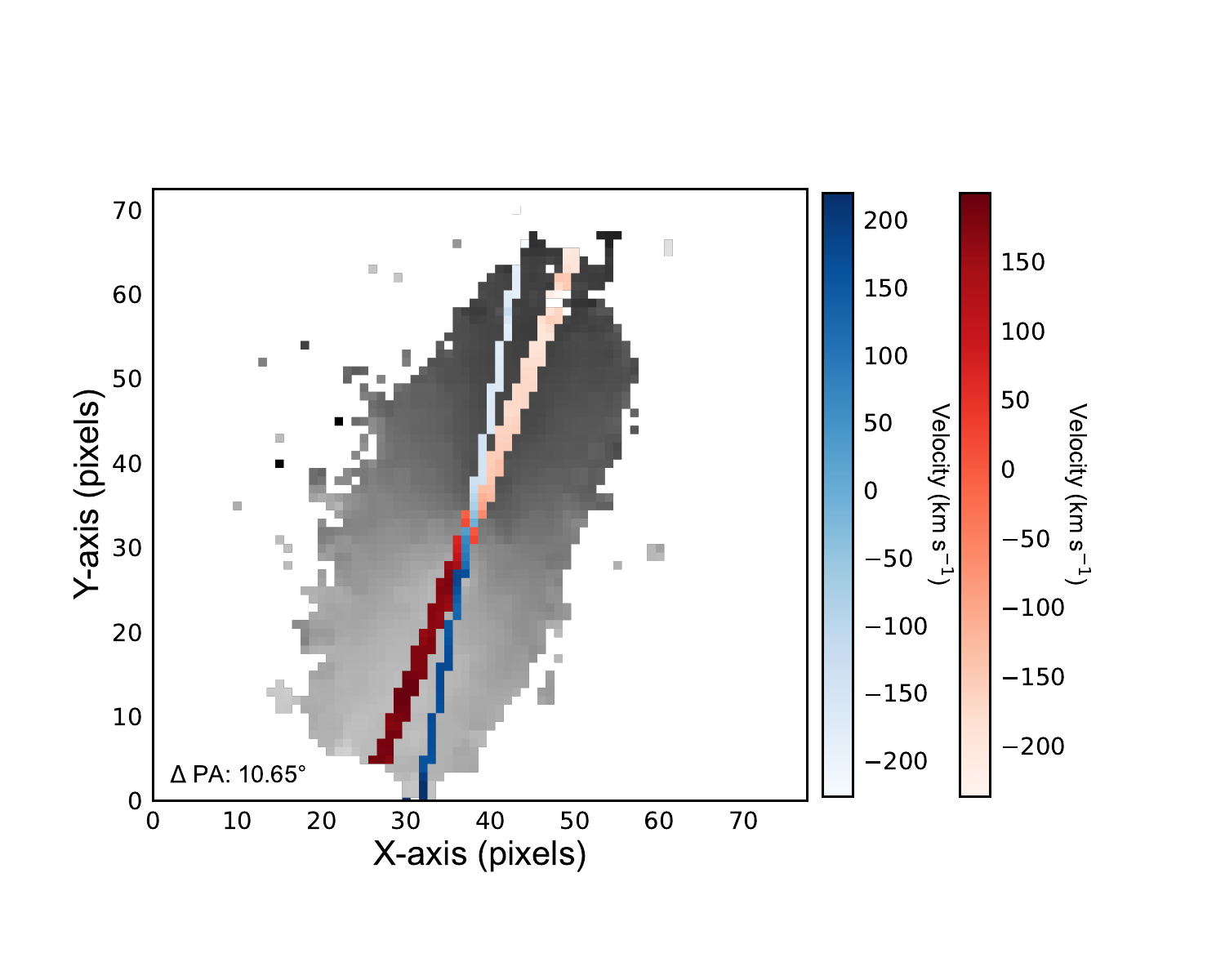}
\caption{Masking the photometric (\texttt{`tractor'}; shown in blue) and kinematic (\texttt{`kinematic'}; shown in red) major axes along their respective PAs on the velocity map for IC\,1199. All masks use the same pixel width $(\pm 3)$; the visually thinner appearance of the Tractor PA is a geometric projection effect due to its steeper orientation relative to the image axes.}
\label{fig:masking_axes}
\end{figure}

To calculate the maximum rotational velocity, we first extract velocity profiles along the galaxy's semi-major axis by constructing pixel masks aligned with both the photometric and kinematic PAs. These masks trace narrow bands of $\pm 3$ pixels along the projected semi-major axis, enabling localized sampling of velocity data. We define a masking function that computes the vertical pixel coordinate $y$ for each horizontal pixel $x$ using the relation:
\begin{equation}
y = x\tan(\theta) + y_c + k
\end{equation}
where $\theta$ is the PA, $y_c$ is the central pixel coordinate, and $k$ spans $\pm 3$ to create a narrow band around the axis. Masking is performed for all three PAs, and each PA is allowed to vary by $\pm 90^\circ$ to ensure alignment with the semi-major axis. The masking results for both the \texttt{`tractor'} derived and \texttt{`kinematic'} PAs are shown in Fig.~\ref{fig:masking_axes}. When plotting the masked velocity values against galactocentric distance, we observe a rise in velocity from the center outward, followed by a plateau beyond a certain point. Based on the distribution of data points, the majority of velocity measurements fall within this plateau region, indicating the galaxy's maximum rotational velocity.

\begin{figure}[t]
\centering
\includegraphics[width=\columnwidth]{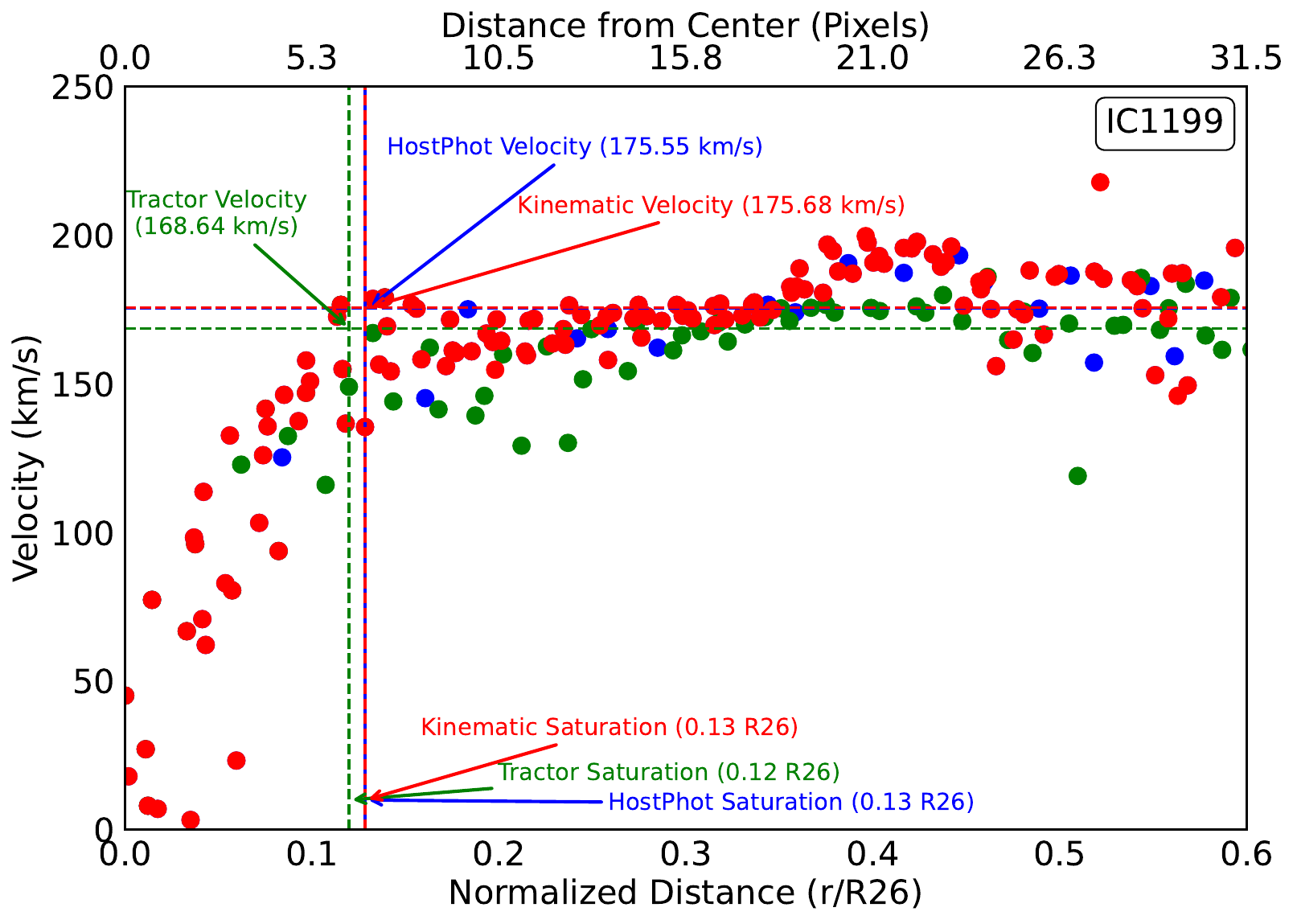}
\caption{Determining the saturation point of the rotational velocity. A line with a slope of 0 is parallel to the x-axis, indicating that the maximum density point is fitted. The vertical line indicates the point where the velocity reaches maximum and then dips. This line is drawn at the point where the maximum value must cross the maximum density line. The fit here is for the \texttt{`kinemetry'}, \texttt{`tractor'}, and \texttt{`HostPhot'} data for IC\,1199.}
\label{fig:saturate}
\end{figure}

We fit a horizontal line at the position of maximum data density, capturing the region of a relatively flat rotation curve. To define the flat portion of the rotation curve, we identify the radial interval where the point-to-point velocity gradient falls below a fixed threshold. Specifically, we require $|\Delta V / \Delta r| < 0.1~{\rm km\,s^{-1}\,pixel^{-1}}$ for MUSE data and $|\Delta V / \Delta r| < 1.0~{\rm km\,s^{-1}\,pixel^{-1}}$ for PMAS data, reflecting the different noise properties and spatial sampling of the two instruments. The first radius satisfying this condition marks the onset of the saturation region used for the horizontal fit shown in Fig.~\ref{fig:saturate}.
This horizontal fit is denoted by the green dashed line for \texttt{`tractor'}, red dashed line for \texttt{`kinemetry'}, and blue dashed line for \texttt{`HostPhot'} in Fig.~\ref{fig:saturate} to show the maximum velocity region. 

\begin{figure}[t]
\centering
\includegraphics[width=\columnwidth]{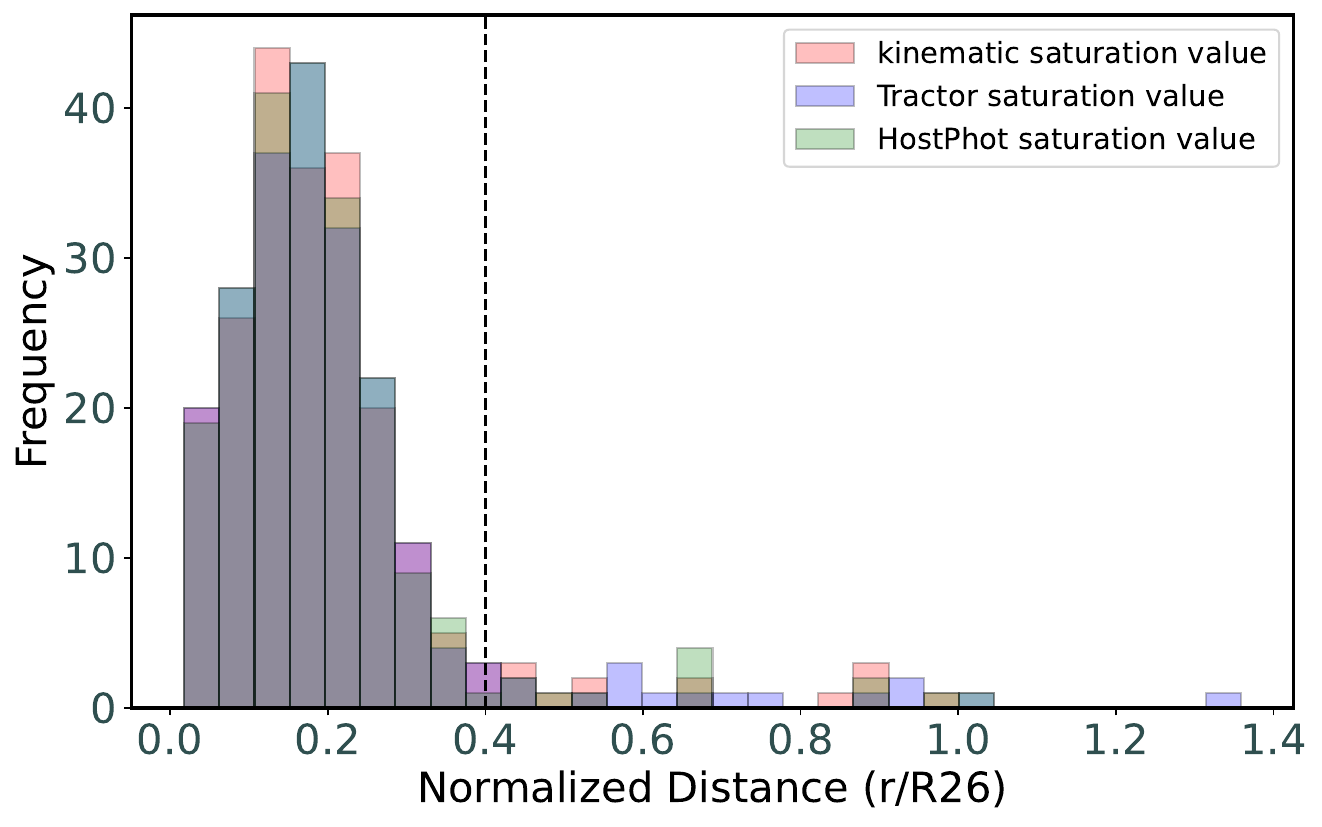}
\caption{Histogram of velocity saturation points. The x-axis is the normalized distance, and the y-axis is the number of galaxies saturated at that corresponding normalized distance. Most of our sample saturates before $0.4\,R_{26}$. A few of the samples do not saturate within $0.4\,R_{26}$ in all three methods.}
\label{fig:saturation_hist}
\end{figure}

We then fit a vertical line to pinpoint the distance where the rotational velocity dips immediately after crossing the maximum data density region as you go from the center towards the edge of the semi-major axis for all \texttt{`tractor'}, \texttt{`HostPhot'}, and \texttt{`kinemetry'} data. The distance corresponding to the first dip in velocity after $y_{\text{max density}}$ is designated as the saturation point, as shown in Fig.~\ref{fig:saturate}. This provides an accurate estimate of the saturation point for most samples. The histogram in Fig.~\ref{fig:saturation_hist} shows that most of our targets saturate below $0.4\,R_{26}$. For the minority of those which saturate beyond $0.4\,R_{26}$, (6.04\% {\sc Kinemetery}, 4.65\% \texttt{`HostPhot'} and 6.98\% \texttt{`tractor'}), the maximum rotational velocity measured at $0.4\,R_{26}$ may underestimate the true maximum rotation velocity, because the rotation curve is still rising at that radius.

Approximately \(80\%\) of the galaxies whose rotation curves continue rising beyond \(0.4\,R_{26}\) are visibly barred in PS1 imaging. This does not mean that the bars themselves extend to \(0.4\,R_{26}\). Instead, barred galaxies exhibit strong non‑circular streaming motions throughout the bar and out to the bar’s corotation radius, which typically lies just beyond the bar end \citep{10.1093/mnras/sty2983}. Inside this bar--corotation region, the velocity field is not yet axisymmetric, so the
azimuthally averaged rotation curve can continue to rise past \(0.4\,R_{26}\). Only outside this region does the curve reach a stable, flat amplitude, explaining why barred galaxies in our sample tend to saturate at larger radii.

\subsection{Calibration and distance estimation} \label{sec:calibration}

To determine reliable TF distances, it is important to calibrate the TF relation specifically for our dataset. While universal calibrations such as those presented by \citet{Kourkchi_2020} offer a general framework, they are often derived from heterogeneous samples with varying photometric systems, velocity definitions, and selection criteria. Applying such calibrations directly to our data risks introducing systematic biases in both the slope and zero point of the TF relation. We derive the observed maximum rotational velocities using three distinct PAs. We also apply an inclination correction to the observed maximum rotational velocity with the respective inclination angle for all three methods before calibration. We measure rotational velocities at a fixed radius of $0.4\,R_{26}$ to match the DESI fiber placement, \citep{2023MNRAS.525.1106S}.  Each method introduces its own set of systematics. These methodological choices necessitate a tailored calibration approach.

The TF slope is sensitive to sample selection, including the inclination angle and range of galaxy distances from the observer included in the sample. A wide distance range introduces distance-dependent selection effects (e.g., magnitude limits and Malmquist bias), which are known to affect the inferred TF slope \citep{Willick_1994}. Unlike previous studies that rely on cluster-based or relative calibrations \citep{2012courtois_calibration, socre_2013, Niel_2014, Kourkchi_2020, Bell_2022, 2023MNRAS.525.1106S, douglass2025desiedrcalibratingtullyfisher}, our sample is not cluster‑selected but instead consists of galaxies distributed across a range of environments, with no common cluster membership. Approximately $10\%$ of our targets have individually measured distances. We identified 20 nearby galaxies with distance estimates anchored by supernova observations, available through the NASA/IPAC Extragalactic Database (NED) and the Pantheon+Ghost compilation \citep{Scolnic_2022, NED}. Because of this, we adopt an absolute calibration approach which directly relates absolute magnitude in the r-band to the corrected rotational velocity \citep{sakai_2000, Tully_2000}, rather than anchoring to a cluster mean or fixed distance scale. Using absolute magnitudes allows us to exploit these direct distance constraints and ensure that the resulting TF zero point is tied to physically meaningful, externally calibrated distances rather than relative offsets. The galaxies with independently measured distances that form our calibration sample are listed in Table~\ref{tab:calibration_sample}.  

We exclude UGC\,2319 due to its extreme luminosity $M = -24.6$ mag. In surveys like DESI, outliers like this are inevitable, but including this galaxy in our ODR fit would affect our kinematic slope to an unphysical scale of -54 $\pm$ 65. A comparison table with and without including this galaxy is given in Appendix \ref{app: sensitivity}. To ensure accurate line-of-sight velocity measurements, we restrict our calibration sample to galaxies with inclination angles greater than $45^\circ$, minimizing projection effects \citep{Tully_2000, Verheijen_2001}. On doing so, we get 13 common galaxies for all three methods, and these are included for calibration for all three methods. The \texttt{`kinemetry'} method has 16 galaxies in total, but does not include MCG-01-33-034, which the other methods include. \texttt{`HostPhot'} excluded UGC\,6332 and NGC\,3811, leading to a final list of 15. \texttt{`tractor'} excluded two more compared to \texttt{`HostPhot'} while having all others the same, leading to 13 final samples for calibration. We perform an ODR fit between $M$ and the logarithm of the corrected rotational velocity using  Eq.~\ref{eq:TF_relation} for each method. 

To calibrate the TF relation for our sample, we first compute the absolute magnitude $M$ for galaxies with known distances using the standard distance modulus relation: 
\begin{equation}
M = m_{R26} - \mu,
\end{equation}
where $m_{R26}$ is the apparent magnitude measured within the $R_{26}$ isophotal radius, and $\mu$ is the distance modulus obtained from independent sources (e.g., NED or Pantheon+Ghost).

\begin{table*}
\centering
\caption{Calibration sample galaxies and their basic properties.}\label{tab:calibration_sample}
\tiny
\resizebox{\textwidth}{!}{
\begin{tabular}{llccccccc}
\toprule
Galaxy name & Supernova  & RA (deg) & Dec (deg) & Distance Modulus & $m_{R26}$ & HostPhot Incl. & Kinematic Incl. & Tractor Incl. \\
\midrule
MCG -01-33-034    & SN\,2006d    & 193.1447 &  -9.7766 & 32.71 & 15.156 & 51.75 & \cellcolor{red!20} 41.12 & 45.59 \\
UGC\,06332   & SN\,2007bc    & 169.8193 &  20.8134 & 34.82 & 12.933 & \cellcolor{red!20} 28.47 & 45.29 & \cellcolor{red!20} 20.82 \\
MCG -02-07-033    & SN\,2008j    &  38.6023 & -10.8438 & 36.08 & 12.862 & 68.10 & 66.04 & 67.84 \\
IC\,1151     &  SN\,1991m   & 239.6348 &  17.4415 & 32.96 & 12.750 & 71.14 & 52.87 & 68.55 \\
IC\,758      &  SN\,1999bg   & 181.0495 &  62.5054 & 32.41 & 13.184 & 50.25 & 64.91 & \cellcolor{red!20} 43.44 \\
MCG+04-42-22 & SN\,2001bf  & 270.3900 &  26.2528 & 33.99 & 13.525 & 48.03 & 54.77 & \cellcolor{red!20} 36.36 \\
MCG-01-09-006 & SN\,2005eq  &  47.2013 &  -7.0406 & 35.33 & 13.477 & 74.95 & 54.62 & 73.57 \\
MCG-01-10-015  &  &  54.6632 &  -5.3474 & 34.77 & 13.797 & 76.86 & 70.19 & 76.66 \\
NGC\,0958    &  SN\,2005a   &  37.6785 &  -2.9390 & 34.46 & 11.889 & 71.45 & 75.28 & 72.51 \\
NGC\,0976    &  SN\,1999dq   &  38.5001 &  20.9768 & 33.39 & 12.083 & \cellcolor{red!20} 38.67 & \cellcolor{red!20} 29.67 & \cellcolor{red!20} 41.32 \\
NGC\,2789    &  SN\,2009cz   & 138.7486 &  29.7303 & 34.86 & 12.575 & \cellcolor{red!20} 38.93 &\cellcolor{red!20}  35.36 &\cellcolor{red!20}  33.23 \\
NGC\,3811    &  SN\,1969c   & 175.3193 &  47.6908 & 33.14 & 12.253 & \cellcolor{red!20} 41.18 & 49.88 &\cellcolor{red!20}  44.46 \\
NGC\,6063    &  SN\,1999ac   & 241.8041 &   7.9790 & 33.53 & 12.950 & 58.45 & 54.89 & 59.15 \\
NGC\,7311    &   SN\,2005kc  & 338.5280 &   5.5699 & 33.82 & 11.793 & 61.50 & 63.80 & 61.09 \\
NGC\,7364    &  SN\,2009fk   & 341.1015 &  -0.1621 & 34.48 & 12.338 & 49.98 & 52.33 & 50.35 \\
UGC\,00139   &  SN\,1998dk   &   3.6327 &  -0.7376 & 33.45 & 13.364 & 60.13 & 65.14 & 66.91 \\
UGC\,02319 $^*$  &      &  42.4733 &  -1.0036 & 38.78 & 14.172 & \cellcolor{red!20} 81.13 & \cellcolor{red!20} 68.29 & \cellcolor{red!20} 81.68 \\
UGC\,04195   & SN\,2000ce    & 121.2786 &  66.7829 & 34.35 & 12.987 & 55.01 & 55.80 & 54.76 \\
UGC\,10710   &   SN\,2000bs  & 256.7188 &  43.1222 & 34.71 & 13.446 & 80.49 & 60.34 & 80.56 \\
UGC\,11758   &  SN\,2006ev   & 322.7401 &  13.9861 & 35.66 & 13.418 & 90.00 & 88.03 & 87.83 \\
\bottomrule
\end{tabular}}

\tablefoot{ Columns: (1) Galaxy name; (2) Supernova name; (3) Right Ascension (RA, J2000, deg);  (4) Declination (Dec, J2000, deg); (5) Distance modulus; (6) Apparent magnitude within the  
isophotal radius $R_{26}$ ($m_{R26}$); (7) HostPhot inclination (deg); (8) Kinematic inclination (deg);  (9) Tractor inclination (deg).  
The galaxy UGC\,02319$^{*}$ is excluded from the calibration fits due to its extreme luminosity, which would bias the regression. For each calibration method, galaxies were selected only if  
their inclination angle exceeded $45^\circ$ as determined by that method, ensuring reliable line-of-sight velocity measurements and minimizing projection effects. Cells highlighted in red indicate inclination values below the $45^\circ$ threshold for that  
method, marking galaxies excluded from the corresponding calibration sample.}
\end{table*}

\begin{figure}[t]
\centering
\includegraphics[width=\columnwidth]{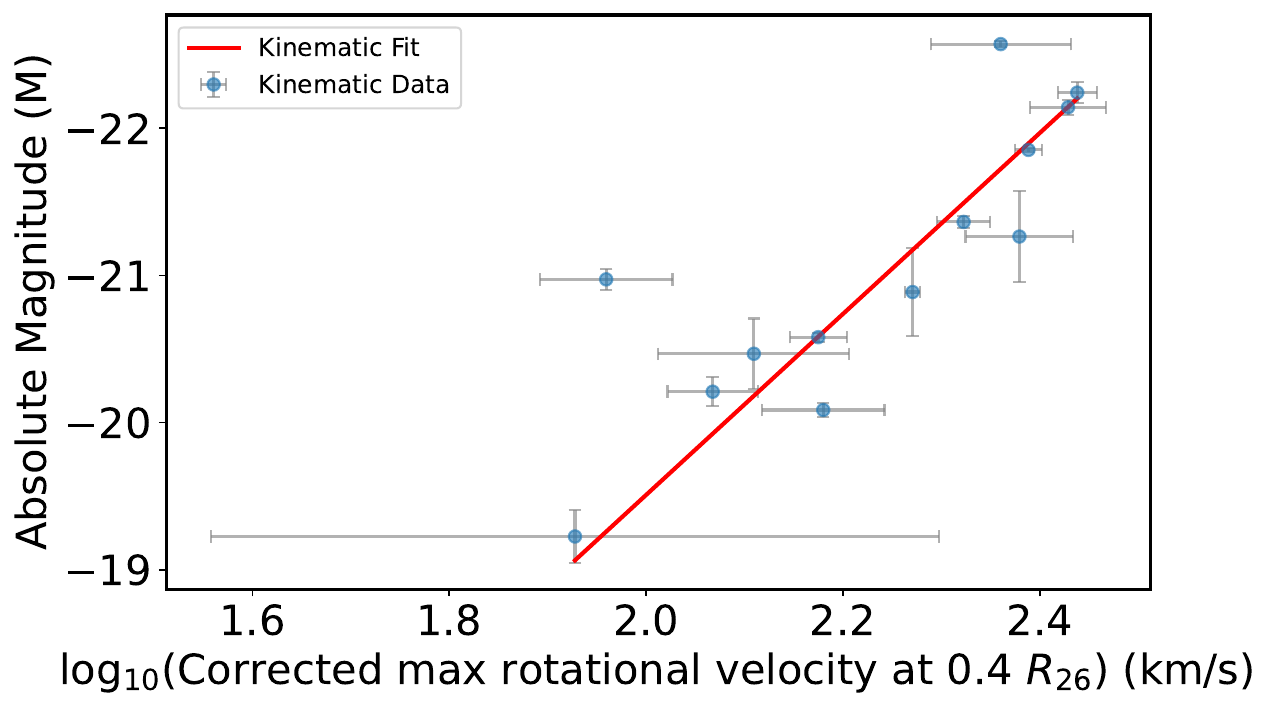}
\includegraphics[width=\columnwidth]{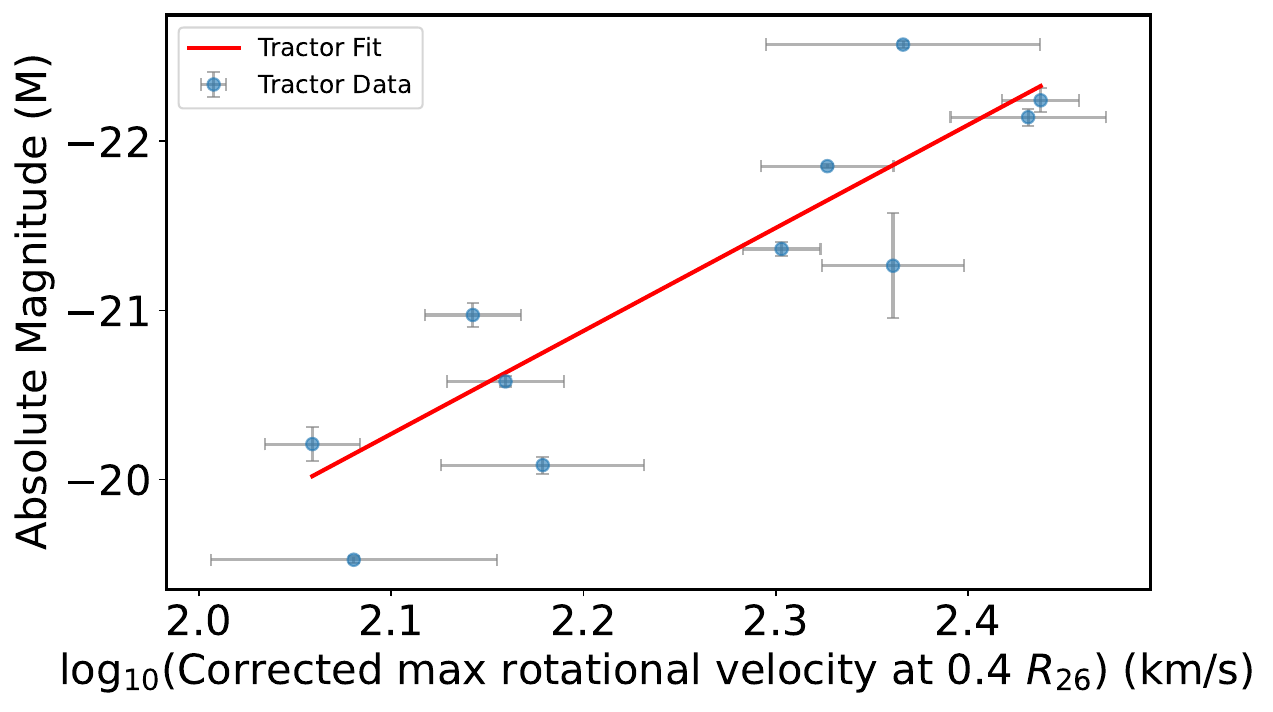}
\includegraphics[width=\columnwidth]{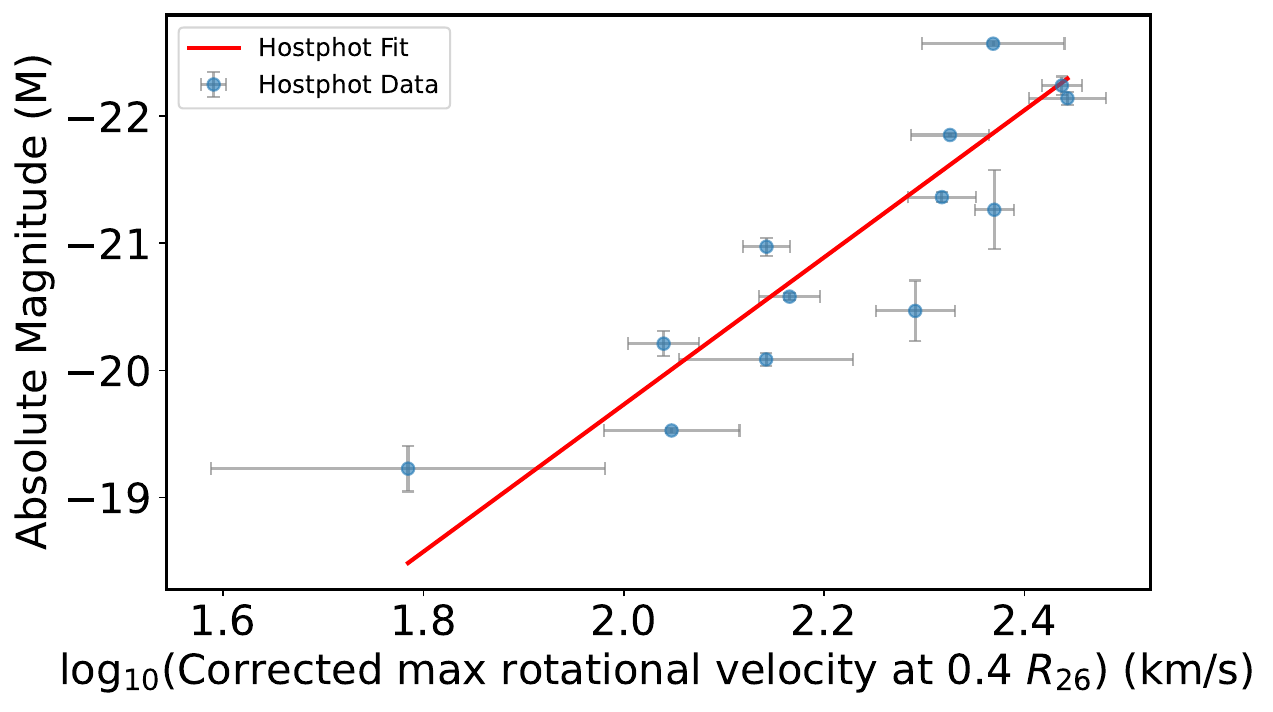}
\caption{Slope fitting for the TF relation using the three rotational–velocity estimates: \texttt{`kinematic'} (top), \texttt{`tractor'} (middle), and \texttt{`HostPhot'} (bottom). In each panel, the slope $a$ is obtained from an orthogonal linear regression, and the x-axis shows the corrected rotational velocity measured at $0.4\,R_{26}$.}
\label{fig:slope_fitting}
\end{figure}

We adopt the Orthogonal Distance Regression (ODR) \citep{Boggs1990} approach rather than ordinary least squares. ODR provides a linear regression framework that minimizes the orthogonal distances between data points and the model line while accounting for uncertainties in both $\log V_{\text{rot}}$ and $M$. This is particularly important for the TF calibration, as both the kinematic and photometric quantities carry significant measurement errors. By treating these uncertainties symmetrically, the resulting slope and zero point are less biased and more physically representative of the intrinsic luminosity--velocity relation. Fig.~\ref{fig:slope_fitting} shows Orthogonal Distance Regression (ODR) fits of absolute magnitude $M$ against the logarithm of the corrected maximum rotational velocity for the three PAs measurement methods, following Eq.~\ref{eq:TF_relation}. For each case, we fit a straight line to derive the TF slope zero point and intrinsic scatter; the resulting parameters are listed and compared with the DESI DR1 TF calibration in Table~\ref{tab:slope_zeropoint}. 

Once calibrated, we apply this relation to the full sample to estimate $M$ for each galaxy and compute the distance modulus and thus distance using Eq.~\ref{eq: mu}.
\begin{equation} \label{eq: mu}
\mu = m_{R26} - M,\\
D = 10^{(\mu + 5)/5}\,\mathrm{Mpc}.
\end{equation}

\begin{table}[t]
\caption{TF calibration slopes and zero points for the three PA determination methods} 
\centering
\resizebox{\columnwidth}{!}{
\begin{tabular}{lccc}
\hline
Method & Slope & Zero point & Intrinsic Scatter \\
\hline
\texttt{`kinemetry'} & $-6.15 \pm 0.88$ & $-22.58 \pm 0.16$ & $0.58 \pm 0.15$ \\
\texttt{`HostPhot'}   & $-5.79 \pm 0.75$ & $-22.63 \pm 0.20$ & $0.49 \pm 0.12$\\
\texttt{`tractor'}    & $-6.09 \pm 0.74$ & $-22.70 \pm 0.20$ & $0.42 \pm 0.07$\\
{\sc DESI PV Survey DR1} & $-7.22 \pm 0.01$ & $various$ &$0.466 \pm 0.001 $\\
\hline
\label{tab:slope_zeropoint}
\end{tabular}}
\tablefoot{The table lists the slopes and zero-point calibrations of the Tully–Fisher (TF) relation derived from maximum corrected rotational velocities using three position–angle (PA) determination
methods: \texttt{kinemetry}, \texttt{HostPhot}, and \texttt{tractor}. For comparison, the DESI-PV Survey calibration \citep{douglass2025desidr1peculiarvelocity} is also included. For the DESI-PV calibration, the slope reflects the luminosity velocity scaling, while the zero point sets the absolute magnitude scale used in their cluster-based TF calibration.}
\end{table}

This approach ensures internal consistency between photometric and kinematic measurements and allows the calibration parameters to be propagated coherently across the full sample. The derived slopes and zero points differ among the PA methods because each technique yields systematically different rotational velocities, producing different best-fit TF parameters. Two additional methodological differences further contribute to the offsets between our results and the DESI-PV calibration: we perform the TF fit directly using absolute magnitudes for both the slope and zero-point determination, whereas \citet{douglass2025desidr1peculiarvelocity} uses a mixed apparent–absolute magnitude calibration.

While our slopes ($\alpha \approx -5.79$ to $-6.15$) are shallower than the canonical values obtained from large TF samples \citep[e.g.][]{douglass2025desidr1peculiarvelocity, Kourkchi_2020}, and the intrinsic scatter of our local calibration is correspondingly higher, this behavior is expected given the small size of the SN‑anchored calibration set. Our aim is not to derive a competitive global TF calibration but to use a homogeneous, SN‑based distance set to compare methods under identical conditions. Imposing a fixed external slope (e.g., the DESI PV DR1 value) would suppress the method‑dependent differences in rotational velocity that we seek to quantify.

\begin{figure*}[t]
\centering
\includegraphics[width=0.48\textwidth]{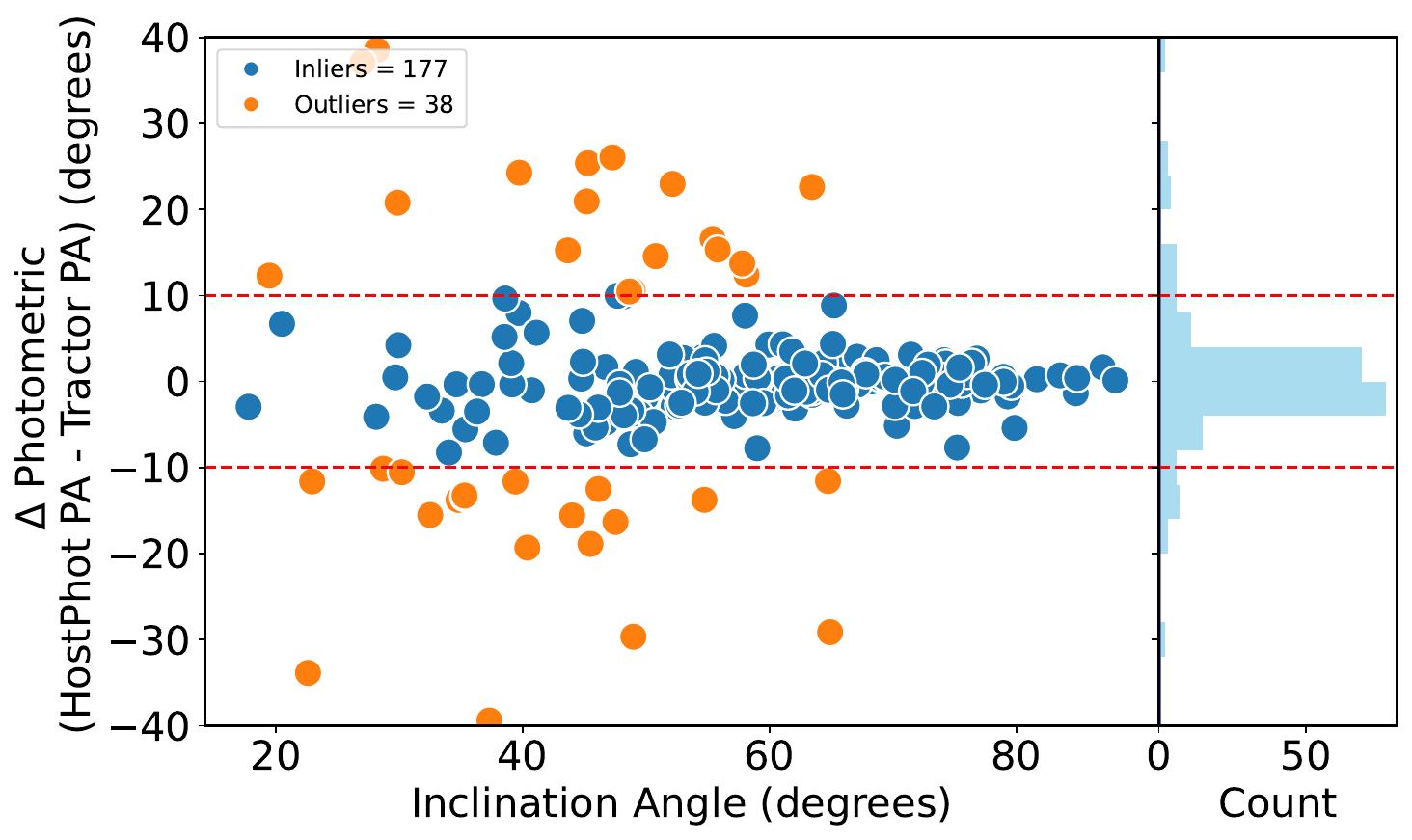}
\includegraphics[width=0.48\textwidth]{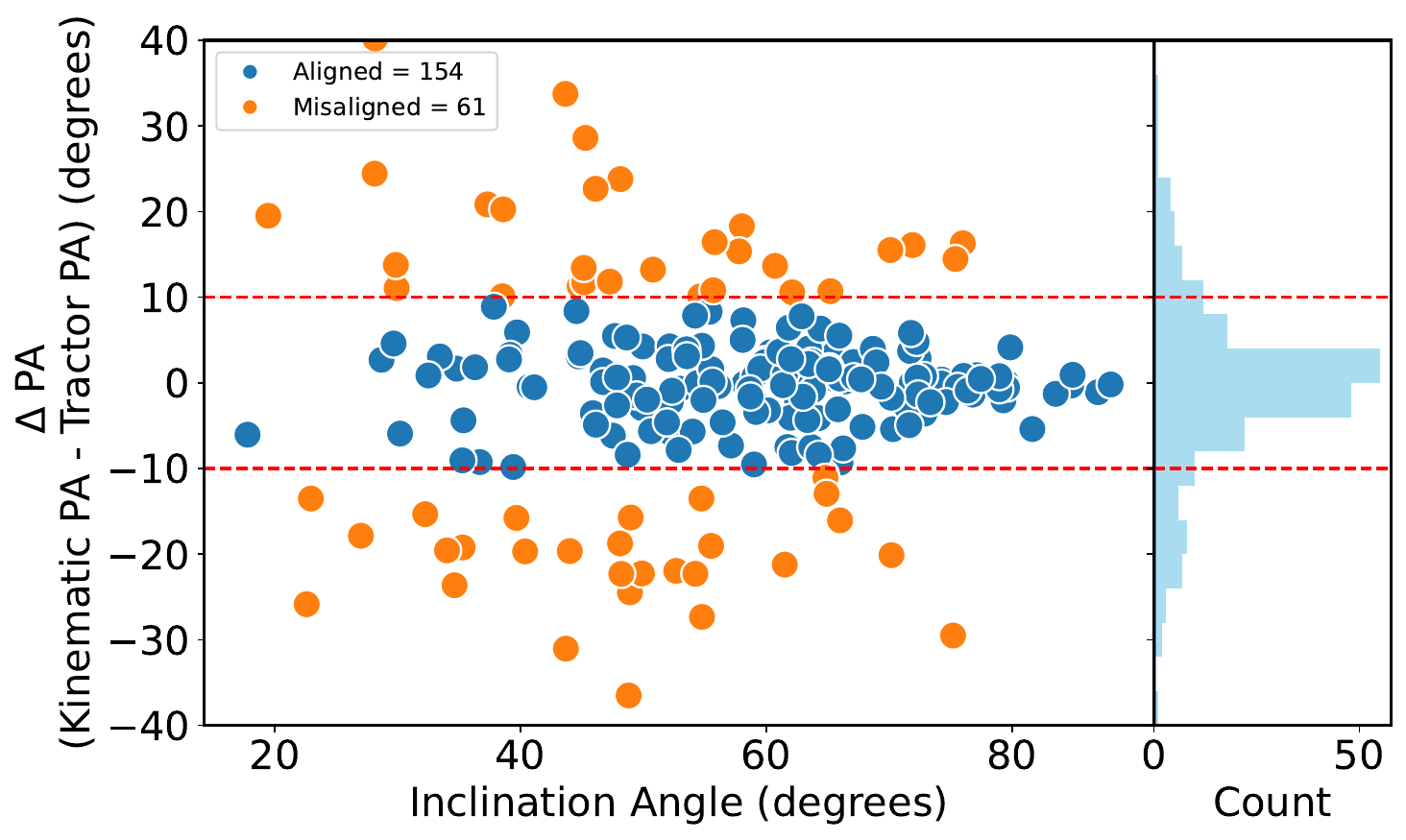}
\caption{Left: Photometric outliers (orange) and inliers (blue) as a function of inclination. 
The apparent reduction in outliers at $i > 70^\circ$ reflects the small number of 
high‑inclination galaxies in the sample. Quantitatively, the scatter in the PA 
difference decreases with inclination (see Sec. \ref{sec:miss}). Right: Difference between the \texttt{`tractor'} photometric PA and the kinematic PA 
(y-axis) plotted against the kinematic inclination angle (x-axis). No systematic trend is visible 
in the PA differences.}
\label{fig:Phot_Vs_Track_Scatter}
\end{figure*}

\section{Results}\label{results}

\subsection{Photometric outliers}\label{sec:outliers}

We find that the photometric PA difference is small for most galaxies: 155 out of 215 (72\%) have $\Delta{\rm PA} < 5^\circ$, 22 galaxies (10\%) lie in the range $5^\circ$--$10^\circ$, and 38 galaxies (18\%) show offsets greater than $10^\circ$. We classify this latter group as photometric outliers. In the left panel of Fig.~\ref{fig:Phot_Vs_Track_Scatter}, we see the outlier distribution in our sample with respect to the kinematic inclination angle $i$. While taking the difference between those PAs, we allow circular rotation $\pm 180\degree$ for \texttt{`HostPhot'} to allow matching the convention between the measurement of two angles. We observe greater scatter at low inclination angles, where the galaxy appears more face-on, consistent with the idea that lower inclination results in unreliable photometry \citep{Skarbinski2022}. Quantitatively, the scatter decreases sharply with inclination: galaxies with $i < 30^\circ$ have a standard deviation of $20.5^\circ$, which drops to $12.1^\circ$ for $30^\circ \le i < 50^\circ$, $7.8^\circ$ for $50^\circ \le i < 70^\circ$, and only $2.2^\circ$ for nearly edge-on systems ($70^\circ \le i < 90^\circ$). 

\subsection{Photometric--kinematic misalignment} \label{sec:miss}

We measure the photometric--kinematic misalignment and find that 118 out of 215 galaxies (55\%) have $\Delta{\rm PA} < 5^\circ$, 36 galaxies (17\%) lie in the range $5^\circ$--$10^\circ$, and 61 galaxies (28\%) show offsets greater than $10^\circ$. 72\% of the sample falls within $0^\circ$--$10^\circ$, which we classify as aligned, while the remaining galaxies are labeled as misaligned.
In the right panel of Fig.~\ref{fig:Phot_Vs_Track_Scatter}, we see the misalignment distribution in our sample with respect to the kinematic inclination angle of the galaxies. We observe a similar pattern to that of photometric outliers, with greater scatter at lower inclination angles and smaller scatter towards higher inclination angles. Galaxies with $i < 30^\circ$ have a standard deviation of $21.7^\circ$, which decreases to $15.6^\circ$ for $30^\circ \le i < 50^\circ$, $9.1^\circ$ for $50^\circ \le i < 70^\circ$, and $7.5^\circ$ for nearly edge-on systems ($70^\circ \le i < 90^\circ$).

\begin{figure}[t]
\centering
\includegraphics[width=9cm]{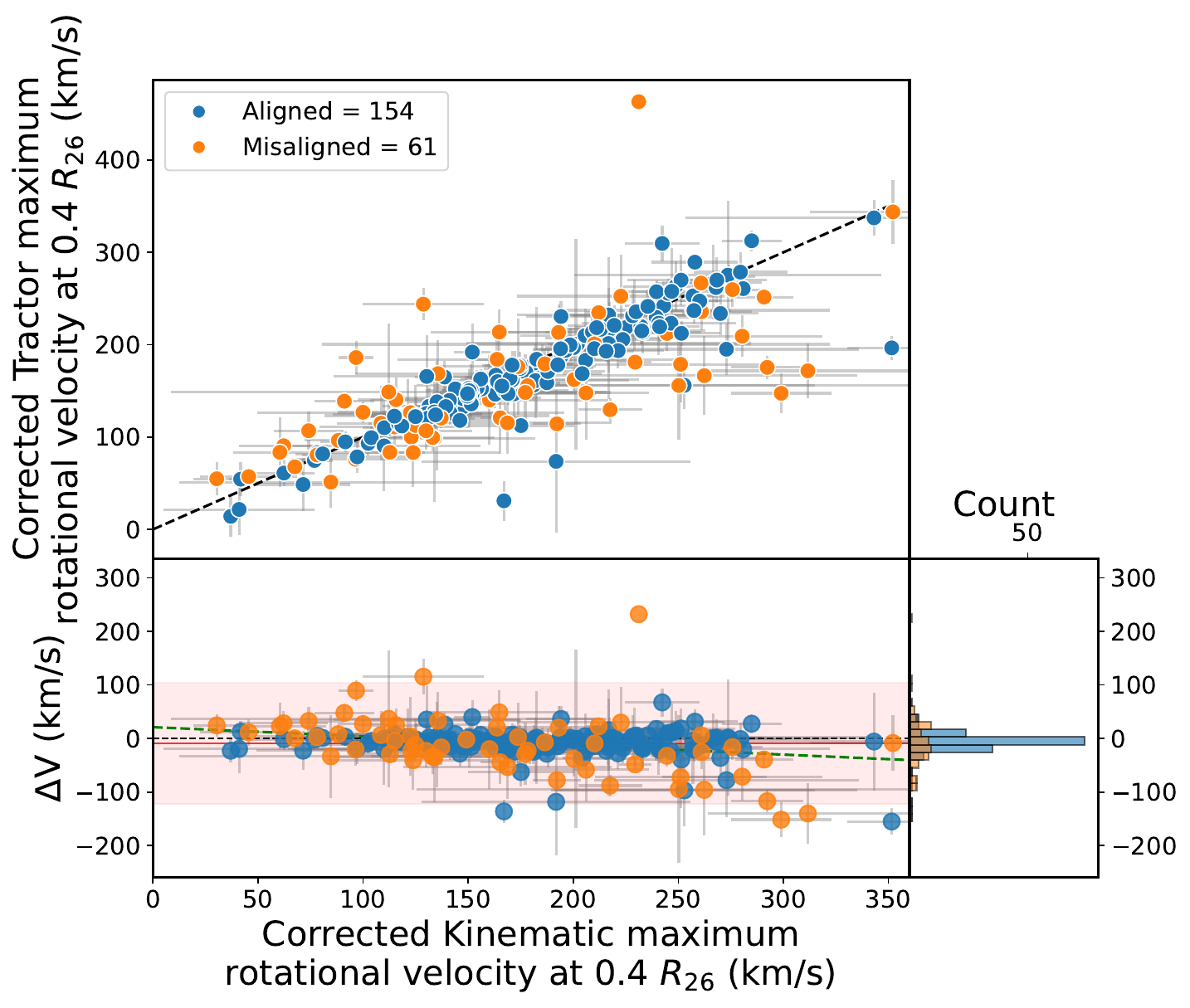}
\caption{Top panel: Direct comparison of the two velocity estimates. Blue points represent aligned galaxies, while orange points are misaligned galaxies, which show larger scatter. Bottom panel: Residual between the two methods. The dashed green line is a linear model fit for the residual. The red solid line is the mean difference, and the black dashed line denotes zero residual. The red shaded band shows the 3$\sigma$ range around the mean residual, indicating the expected variation due to random scatter. Galaxies outside this band are significant outliers in the velocity comparison. Right panel: Histogram showing that the misaligned galaxies are more scattered than the aligned galaxies.}
\label{fig: velocity_not_corrected}
\end{figure}

We compare the corrected maximum rotational velocities derived using two methods. Fig.~\ref{fig: velocity_not_corrected} (top panel) plots the \texttt{`kinemetry'} velocities on the abscissa and the \texttt{`tractor'} velocities on the ordinate, while the bottom panel shows the difference between the two measurements. Misaligned galaxies exhibit larger discrepancies than aligned galaxies, with a scatter more than twice as large (SD $=58.3$ vs.\ $25.5\ \mathrm{km\ s^{-1}}$) and a larger mean offset ($-12.1$ vs.\ $-8.3\ \mathrm{km\ s^{-1}}$). While \cite{2023MNRAS.525.1106S} reported an average difference between the maximum rotational velocity from MaNGA DR15 observations \citep{Bundy_2015} with \texttt{`tractor'} estimate to be 25 $\mathrm{km\ s^{-1}}$, we find the overall mean difference to be $\Delta V = -9.36\ \mathrm{km\ s^{-1}} \, \pm 2.56 \, \mathrm{km\ s^{-1}}$. The smaller velocity offset found here compared to MaNGA is due to the higher spatial resolution of AMUSING and, to a lesser extent, PISCO. MaNGA's coarser resolution leads to significantly stronger beam smearing, which suppresses the observed velocity gradient and can artificially amplify Photometric--Kinematic PA misalignment. 

The mean error between the methods is about 3.5\%. When splitting the sample by alignment, aligned galaxies have a mean difference of $-8.29\ \mathrm{km\ s^{-1}}$ with standard deviation $\mathrm{SD}=25.45\ \mathrm{km\ s^{-1}}$, whereas misaligned galaxies show a mean difference of $-12.06\ \mathrm{km\ s^{-1}}$ with $\mathrm{SD}=58.30\ \mathrm{km\ s^{-1}}$. This is also shown by the histogram plot in Fig.~\ref{fig: velocity_not_corrected}, highlighting that the misaligned galaxies are more spread out compared to the aligned. These results indicate that misaligned systems not only produce a larger systematic offset between methods but also a substantially greater scatter, implying that differences in PA lead to both biased and more uncertain estimates of the maximum rotational velocity.

\begin{figure}[t]
\centering
\includegraphics[width=9cm]{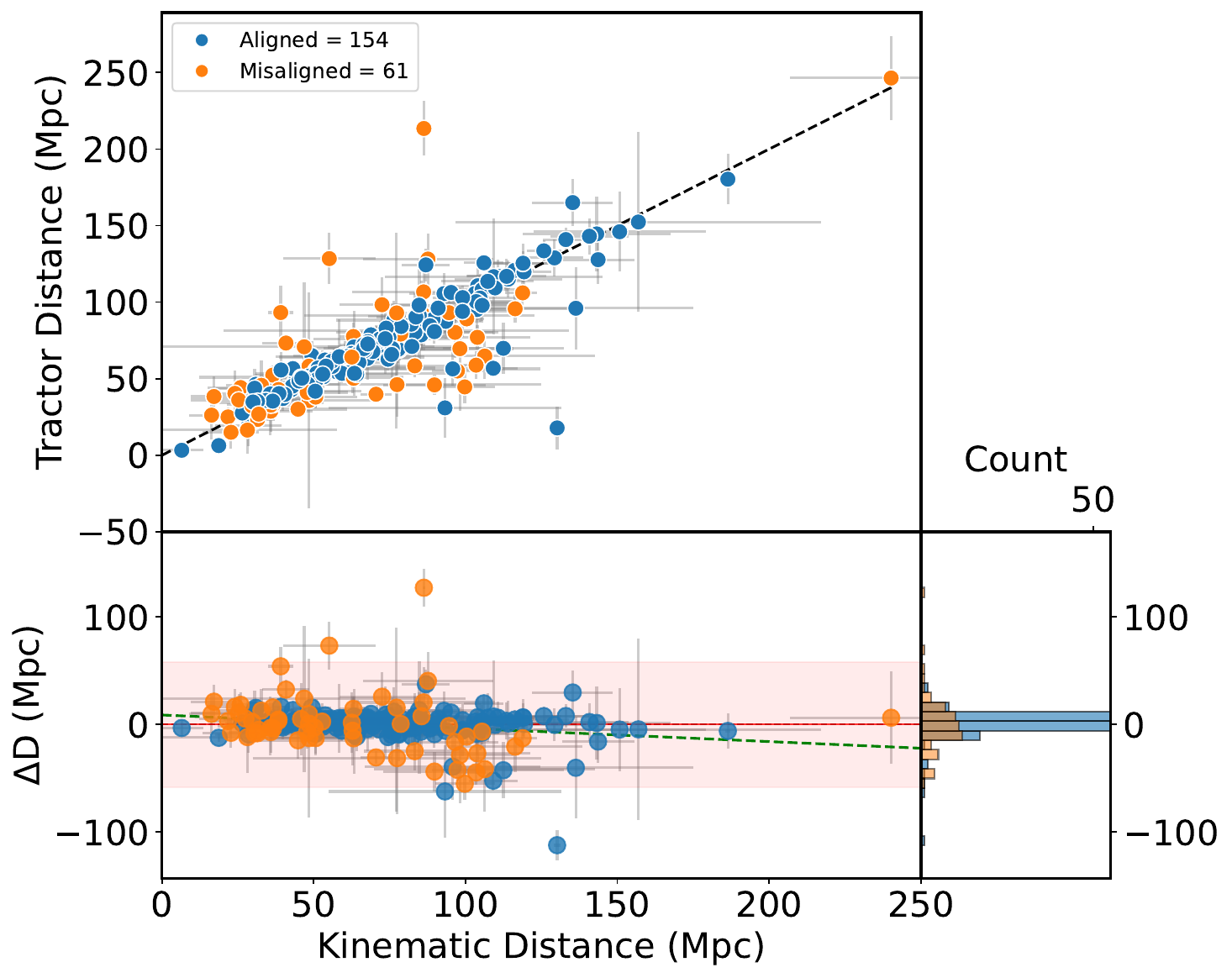}
\caption{Top panel: Direct comparison of the two distance estimates from the \texttt{`tractor'} and \texttt{`kinemetry'}. Bottom panel: Residual between the two methods. The dashed green line is a linear model fit for the residual. The red solid line is the mean difference, and the black dashed line (overlapped with the red solid) denotes the zero residual. The green dashed line crosses the black dashed line at $\sim 72\,\mathrm{Mpc}$, highlighting that the method overestimates distances for nearby galaxies but underestimates distances for distant galaxies. The error bars are higher for misaligned and distant galaxies. The solid red line is the mean residual, and shaded red indicates the $3\sigma$ range, representing statistical variation in distance discrepancies. Right panel: The histogram highlights that the misaligned galaxies show higher scatter than aligned galaxies.}
\label{fig: systematic_difference}
\end{figure}

\subsection{Distance estimation and systematic correction}\label{sec:rescalibration}

We estimate distances using all three methods, with the results of 15 randomly selected galaxies presented in Appendix \ref{app: appendix_distance}. To assess the systematic uncertainty in DESI-TF distances, we perform both residual and Bland-Altman analyses and calculate the Mean Percentage Error (MPE) for the distance estimates.

Fig.~\ref{fig: systematic_difference} presents a comparison between the \texttt{kinemetry} and \texttt{Tractor} distance estimates. The lower panel displays the residuals, defined as $\Delta D = D_{\rm Tractor} - D_{\rm Kinematic}$ in Mpc. Misaligned galaxies exhibit a notably larger spread in their distance estimates compared to the aligned galaxies, showing roughly twice the scatter (SD $\approx 28\,\mathrm{Mpc}$) relative to the aligned sample (SD $\approx 15\,\mathrm{Mpc}$), consistent with the pattern observed in Fig.~\ref{fig: velocity_not_corrected}. This difference in scatter is substantial and statistically significant.

While most of the aligned galaxies show less scatter, two galaxies appear as outliers with larger discrepancies between their photometric and kinematic distance estimates. Upon closer examination, we find that these systems are highly inclined, and the inclinations derived from photometric and kinematic methods differ by approximately $20 \degree$ (see Fig.~\ref{app_fig:ellipse}). A discrepancy of $\sim 20^\circ$ corresponds to a large difference in the implied axis ratio, and this directly affects the deprojection correction $V_{\rm rot} = V_{ obs}/\sin i$. A smaller inclination (more face-on) produces a larger correction and therefore a larger TF distance, while a larger inclination yields a smaller correction and a closer TF distance. Although their PAs are quite similar (within $10^\circ$ ), the velocity measured along the photometric semi-major axis can differ significantly from that along the kinematic axis, leading to systematically underestimated distances. 

To assess whether inclination or PA misalignment dominates the error budget, we estimated the fractional velocity error using $\Delta V/V = \Delta i / \tan i$ for inclination and $\Delta V/V = 1 - \cos(\Delta{\rm PA})$ for PA offsets. For typical values in our sample, PA uncertainties contribute only $\sim 0.2\%$ to the velocity error, whereas inclination uncertainties contribute $\sim 8\%$. Thus, inclination errors overwhelmingly dominate the geometric contribution to the TF distance scatter, even in systems where the photometric and kinematic PAs are well aligned. This is consistent with the findings from \cite{2023MNRAS.525.1106S} that rotational velocity difference amplifies for highly inclined galaxies.

Statistically, we find a mean bias of \(\Delta D = -0.07\,\mathrm{Mpc}\) driven by a small number of large negative residuals, and a median bias of \(\Delta D = +1.98\,\mathrm{Mpc}\) indicating that most galaxies exhibit a slight positive offset. However, this bias is not uniform across the sample: Tractor distances tend to exceed the kinematic distance for galaxies less than $\sim 72 \,\mathrm{Mpc}$ and become systematically lower at larger distances. This distance-dependent sign flip does not imply that PA misalignment is larger for nearby galaxies.  

\begin{figure}[t]
\centering
\includegraphics[width=9cm]{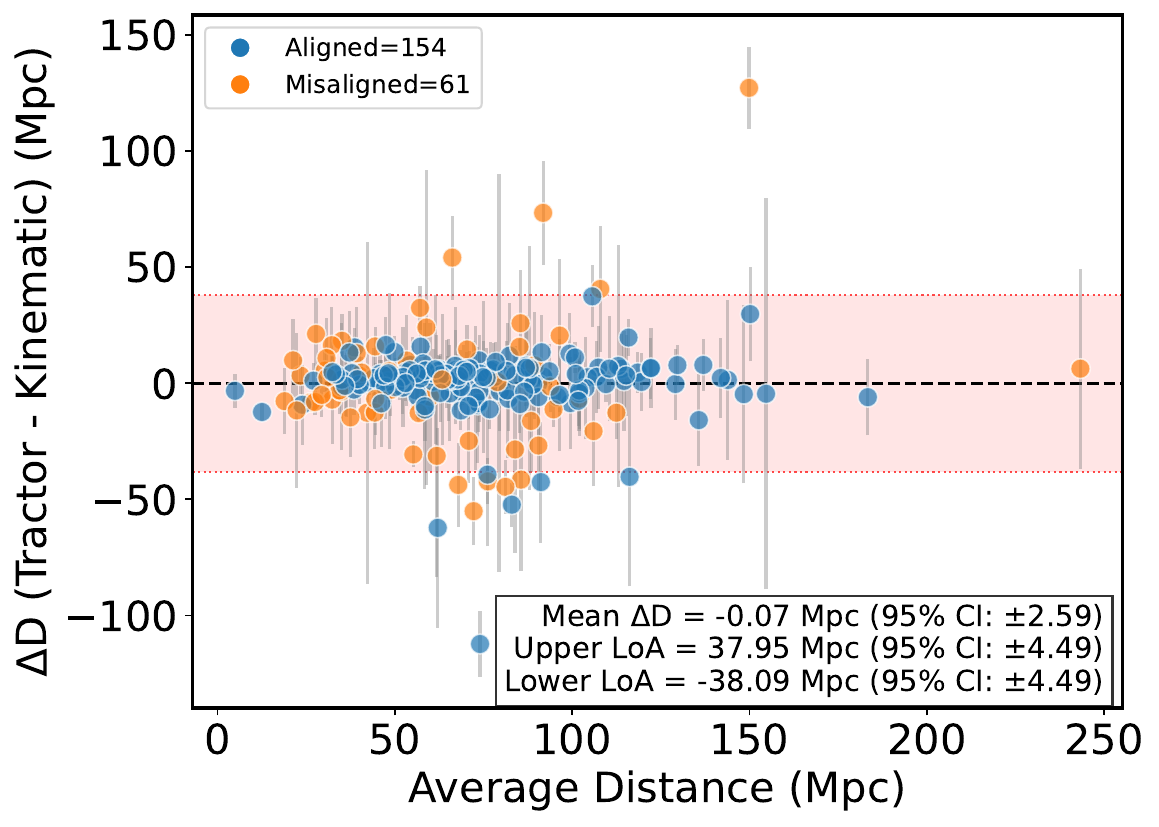}
\caption{Bland-Altman Plot for distance estimation using \texttt{`kinemetry'} for kinematic and \texttt{`tractor'} for DESI projection. Blue-colored dots represent aligned galaxies, while orange-colored dots represent misaligned samples. Misaligned galaxies spread more compared to aligned ones. The error bar is higher for outliers and distant galaxies. The dashed dark line represents the mean bias and the red band represents 95$\%$ confidence interval for limits of agreement (LoA); (mean difference $\pm $1.96 $\times$ $\sigma$).}
\label{fig: bland_altman}
\end{figure}

We find that aligned galaxies have a negative mean offset of $\Delta D = -0.26\,\mathrm{Mpc}$ with a standard deviation of $14.52\,\mathrm{Mpc}$, while misaligned galaxies show a positive mean offset of $\Delta D = 0.41\,\mathrm{Mpc}$ and a larger standard deviation of $28.07\,\mathrm{Mpc}$. This indicates that misalignment not only increases the systematic bias but also amplifies the scatter in distance estimates. Figures~\ref{fig: systematic_difference}--\ref{fig: bland_altman} further show that misaligned galaxies (orange points) consistently exhibit larger scatter than aligned galaxies (blue points) across the full distance range, reinforcing that photometric--kinematic PA misalignment is the primary contributor to increased TF distance dispersion.

The Bland-Altman analysis in Fig.~\ref{fig: bland_altman} reveals several important characteristics of our measurement comparison. The clustering of points around the mean difference line confirms the negligible negative mean bias observed in our residual analysis. However, the wide limits of agreement (spanning approximately $-38.09\,\mathrm{Mpc}$ to $+37.95 \,\mathrm{Mpc}$ [95\%  CIs $\pm 4.49$]) highlights the substantial individual variability between methods. This shows nearly double the scatter for misaligned galaxies in comparison to aligned galaxies. 

These results underscore that photometric--kinematic misalignments introduce non-negligible scatter in DESI-TF distances and highlight the need for robust bias correction in DESI-derived TF distances. This is also true when we compare method-to-method discrepancies in all three \texttt{`HostPhot'}, \texttt{`tractor'}, and \texttt{`kinemetry'} comparisons. See Appendix \ref{App: method to method comparison}.

We find the Mean Percentage Error for the full sample to be $2.8\%\pm1.96\%$ (SE), with aligned galaxies exhibiting a lower MPE of $0.8\%$ and misaligned galaxies showing a substantially larger MPE of $7.9\%$. These values motivate three practical uncertainty regimes: well-aligned systems ($\Delta{\rm PA}<10^\circ$) are consistent with a $\sim 1\%$ fractional uncertainty (as a $\sigma$-equivalent term), the full-sample MPE corresponds to a global $\sim 3.5\%$ $\sigma$-equivalent uncertainty, and strongly misaligned or kinematically incomplete galaxies require an upper threshold of $\sim 10\%$ (also expressed as a $\sigma$-equivalent term).

As an immediate and conservative step, we recommend DESI combine this additional uncertainty in quadrature with the published \texttt{Tractor} errors. The combined uncertainty is:
\begin{equation}
\sigma_{\rm sym}=\sqrt{\sigma_{\rm Tractor}^2 + \left(0.035\,D_{\rm Tractor}\right)^2},
\end{equation}
where $\sigma_{\rm sym}$ is the total symmetric uncertainty, $\sigma_{\rm Tractor}$ is the published \texttt{Tractor} distance uncertainty from DESI, and $D_{\rm Tractor}$ is the corresponding \texttt{Tractor}-based distance estimate. For galaxies with well-constrained kinematic PAs and photometric--kinematic alignment ($\Delta{\rm PA}<10^\circ$), the quadrature term should instead adopt the $\sim 1\%$ $\sigma$-equivalent uncertainty. Conversely, for kinematically known but strongly misaligned or incomplete systems, the quadrature term should use the upper $\sim 10\%$ $\sigma$-equivalent uncertainty.

We also quantify the residual trend shown in Fig.~\ref{fig: systematic_difference} by fitting a simple linear model,
\begin{equation}
\Delta D = D_{\rm Tractor} - D_{\rm Kinematic} = a + b\,D_{\rm Tractor},
\end{equation}
where the intercept is $a = -14.36 \pm 2.85\,\mathrm{Mpc}$ and the slope is $b = 0.199 \pm 0.036$. The slope is highly significant ($p = 7.87 \times 10^{-8}$), and the bias crosses zero near $D_{\rm Tractor} \simeq 72\,\mathrm{Mpc}$, indicating that \texttt{Tractor} distances are overestimated at small distances and underestimated at large distances. The corrected distance for an individual galaxy can then be obtained as
\begin{equation}
D_{\rm corr} = D_{\rm Tractor} - \Delta D.
\end{equation}
Here, $\Delta D$ represents the bias estimated from the residuals, which can be positive or negative depending on distance; subtracting it automatically corrects over- or underestimation. This linear correction addresses the systematic distance-dependent bias and is valid within the redshift range of our calibration sample ($z \lesssim 0.04$), complementing the quadrature uncertainty prescription that accounts for the random scatter driven by PA misalignment and morphological complexity. However, the linear fit does not eliminate the heteroskedastic component of the uncertainty: galaxies that are strongly misaligned, barred, or morphologically disturbed retain substantially larger residual scatter even after the bias correction. We therefore caution that the linear correction should not be used as a high-precision adjustment for such systems, but rather as a first-order debiasing step that must still be accompanied by the appropriate $\sigma$-equivalent uncertainty term (1\%, 3.5\%, or 10\%, depending on alignment and kinematic completeness).

\section{Conclusions and future work}\label{conclusion}

We analyze a sample of 215 galaxies to quantify systematic offsets between DESI TF distances and kinematic distances derived from PMAS and MUSE. For consistency with DESI, we evaluated $V_{\max}$ at $0.4\,R_{26}$. We find that 11–15 galaxies $\sim (5-7)\% $of the sample do not reach velocity saturation by this radius, with $\approx 80 \%$ of these showing the presence of a bar, potentially affecting the saturation radius. Their TF distances are likely slightly underestimated. This represents a modest contribution to the overall scatter, which is captured in the MPE of 2.8\% (corresponding to a standard-deviation equivalent of $\approx$3.5\%). 

Approximately 28\% of our sample exhibits photometric--kinematic PA misalignment ($\Delta{\rm PA}>10^\circ$), and while the median distance bias between \texttt{Tractor} and kinematic estimates is small ($\sim 2\,\mathrm{Mpc}$), the object-to-object scatter is large and strongly alignment-dependent. Aligned galaxies show a dispersion of $\sim 15\,\mathrm{Mpc}$ and an MPE of only 0.8\%, whereas misaligned systems display twice the scatter ($\sim 28\,\mathrm{Mpc}$) and a substantially larger MPE of 7.9\%. This demonstrates that random scatter driven by PA misalignment and morphological disturbance, rather than a coherent systematic offset, is the dominant source of uncertainty in DESI-TF distances. This increased object-to-object variance propagates directly into PV uncertainties, since velocities scale as $v_{\rm pec} = H_0 \Delta D$. As a result, PA-induced systematics primarily contribute an additional stochastic term to the velocity covariance rather than a coherent shift in the inferred distance scale. These effects, therefore, increase the effective noise in PV analyses and may contribute to increased uncertainties in derived large-scale velocity statistics. A detailed forward modeling of this contribution within a full survey-level analysis is left for future work.

We identify a mild distance-dependent bias and provide a linear correction valid within our calibration range ($z\lesssim 0.04$), but this adjustment removes only the mean trend and does not reduce the heteroskedastic residuals. Consequently, the primary recommendation is to adopt a $\sigma$-equivalent fractional uncertainty of 3.5\% for DESI-TF distances, refined to 1\% for well-aligned galaxies and increased to 10\% for strongly misaligned, barred, or morphologically disturbed systems. This combined treatment improves the robustness of DESI-based TF distances for cosmological applications without altering the underlying TF calibration.

Future work can extend this analysis by exploring additional parameters that define the kinematics of galaxies, such as their morphological and dynamical characteristics. For instance, it would be valuable to differentiate the systematic bias across distinct galaxy types, such as fast rotators versus slow rotators, similar to the misalignment studies presented in \citep{Ene_2018}. A detailed investigation into how these biases evolve across different galaxy populations could offer deeper insights into the nature of projection and calibration errors, ultimately leading to more refined correction strategies. Additionally, a focused analysis of scatter-dominated targets and outliers, some of which show significant variability between photometric and kinematic methods regardless of PA alignment, could help isolate secondary contributors to the observed residuals. While our analysis does not explicitly model selection effects, future work could investigate whether classical biases such as Eddington or Malmquist bias contribute to the observed scatter in TF residuals, particularly in magnitude-limited samples where luminosity thresholds and intrinsic scatter may interact with survey depth \citep{Teerikorpi1997, Willick_1997}.

\section*{Data availability}

The associated data and code are available on Zenodo via DOI:
\href{https://doi.org/10.5281/zenodo.17925385}{10.5281/zenodo.17925385}.

\begin{acknowledgements}
The authors gratefully acknowledge Rajendra Adhikari for his contributions during the thesis stage of this work.
U.S. and L.G. acknowledge financial support from the Spanish Ministerio de Ciencia e Innovaci\'on (MCIN) and the Agencia Estatal de Investigaci\'on (AEI) 10.13039/501100011033 under the PID2023-151307NB-I00 SNNEXT project, from Centro Superior de Investigaciones Cient\'ificas (CSIC) under the PIE project 20215AT016 and the program Unidad de Excelencia Mar\'ia de Maeztu CEX2020-001058-M, and from the Departament de Recerca i Universitats de la Generalitat de Catalunya through the 2021-SGR-01270 grant.
This material is based upon work supported by the U.S. Department of Energy (DOE), Office of Science, Office of High-Energy Physics, under Contract No. DE–AC02–05CH11231, and by the National Energy Research Scientific Computing Center, a DOE Office of Science User Facility under the same contract. Additional support for DESI was provided by the U.S. National Science Foundation (NSF), Division of Astronomical Sciences under Contract No. AST-0950945 to the NSF’s National Optical-Infrared Astronomy Research Laboratory; the Science and Technology Facilities Council of the United Kingdom; the Gordon and Betty Moore Foundation; the Heising-Simons Foundation; the French Alternative Energies and Atomic Energy Commission (CEA); the National Council of Humanities, Science and Technology of Mexico (CONAHCYT); the Ministry of Science, Innovation and Universities of Spain (MICIU/AEI/10.13039/501100011033), and by the DESI Member Institutions: \url{https://www.desi.lbl.gov/collaborating-institutions}. 
The DESI Legacy Imaging Surveys consist of three individual and complementary projects: the Dark Energy Camera Legacy Survey (DECaLS), the Beijing-Arizona Sky Survey (BASS), and the Mayall z-band Legacy Survey (MzLS). DECaLS, BASS, and MzLS together include data obtained, respectively, at the Blanco telescope, Cerro Tololo Inter-American Observatory, NSF’s NOIRLab; the Bok telescope, Steward Observatory, University of Arizona; and the Mayall telescope, Kitt Peak National Observatory, NOIRLab. NOIRLab is operated by the Association of Universities for Research in Astronomy (AURA) under a cooperative agreement with the National Science Foundation. Pipeline processing and analyses of the data were supported by NOIRLab and the Lawrence Berkeley National Laboratory. Legacy Surveys also uses data products from the Near-Earth Object Wide-field Infrared Survey Explorer (NEOWISE), a project of the Jet Propulsion Laboratory/California Institute of Technology, funded by the National Aeronautics and Space Administration. Legacy Surveys was supported by: the Director, Office of Science, Office of High Energy Physics of the U.S. Department of Energy; the National Energy Research Scientific Computing Center, a DOE Office of Science User Facility; the U.S. National Science Foundation, Division of Astronomical Sciences; the National Astronomical Observatories of China, the Chinese Academy of Sciences, and the Chinese National Natural Science Foundation. LBNL is managed by the Regents of the University of California under contract to the U.S. Department of Energy. The complete acknowledgments can be found at \url{https://www.legacysurvey.org/}.
Any opinions, findings, and conclusions or recommendations expressed in this material are those of the author(s) and do not necessarily reflect the views of the U. S. National Science Foundation, the U. S. Department of Energy, or any of the listed funding agencies.
The authors are honored to be permitted to conduct scientific research on I'oligam Du'ag (Kitt Peak), a mountain with particular significance to the Tohono O’odham Nation. 
\end{acknowledgements}

\bibliographystyle{aa}
\bibliography{tf}

\begin{appendix}
    
\onecolumn

\section{Method to method comparison} \label{App: method to method comparison}

\begin{figure*}[h]
\centering
\includegraphics[width=0.48\textwidth]{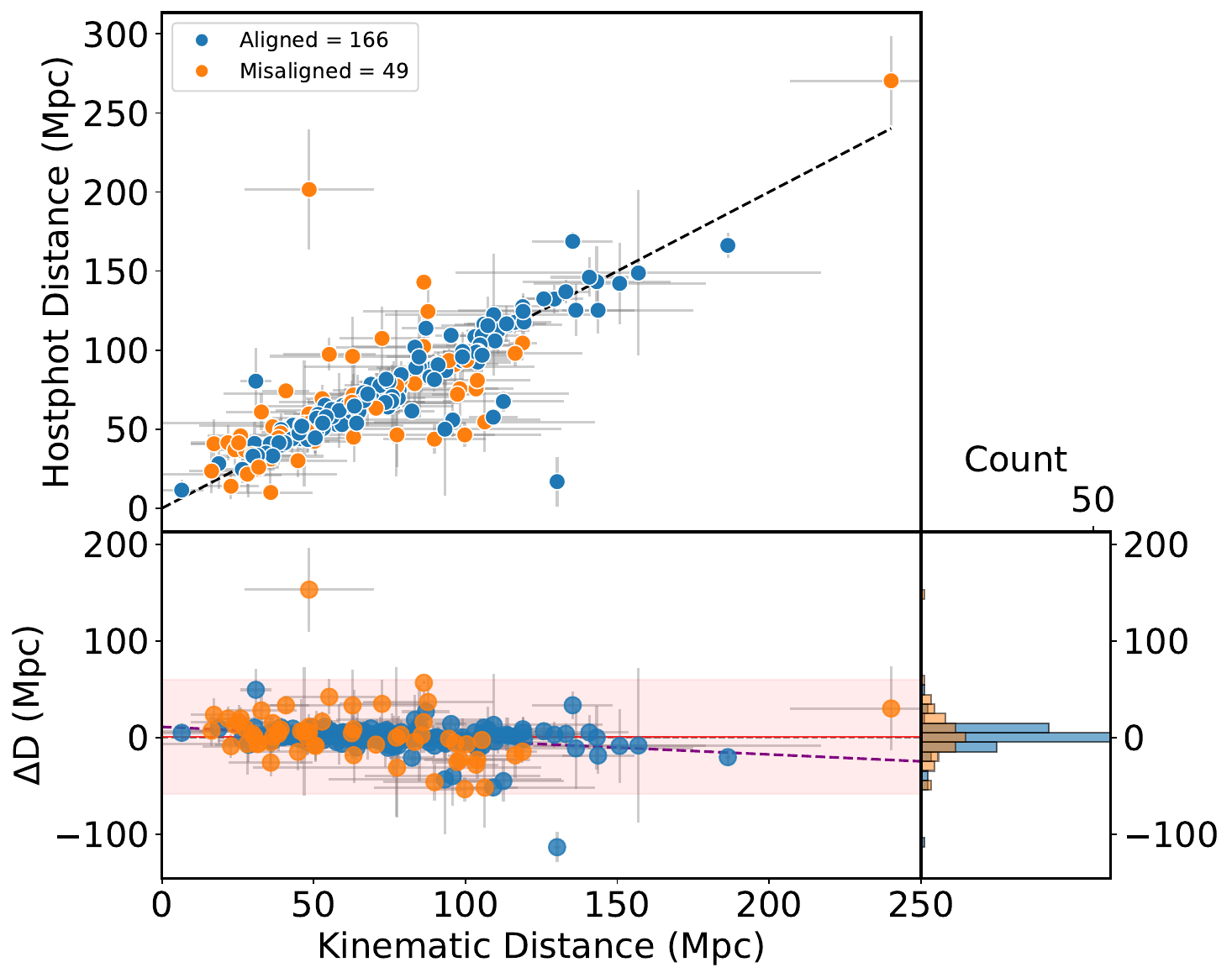}
\includegraphics[width=0.48\textwidth]{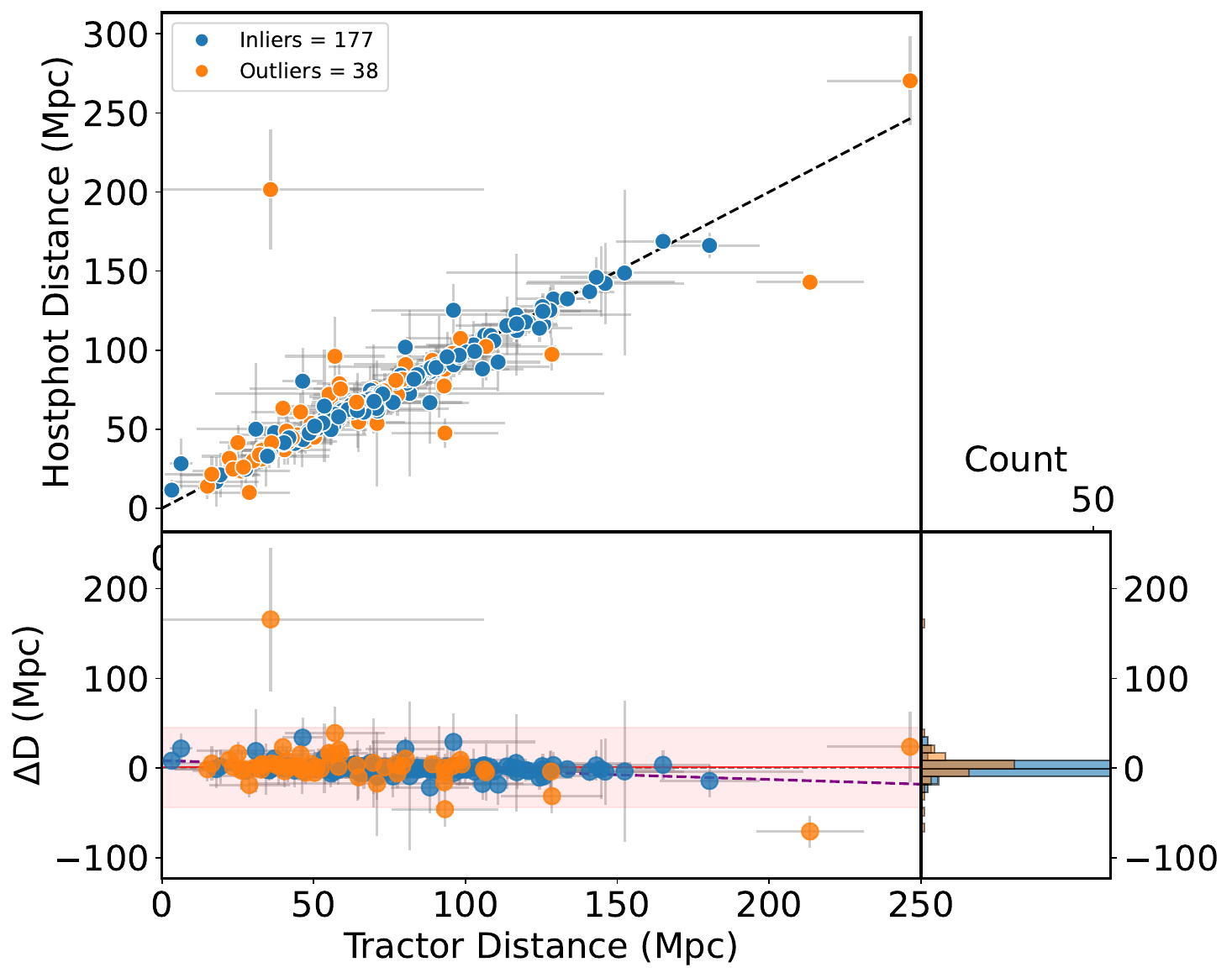}
\caption{Residual analysis of TF distance estimates. 
Left: Comparison between HostPhot and Kinematic distances. 
Right: Comparison between HostPhot and Tractor distances. 
In each panel, the top shows the direct distance comparison with error bars, the bottom shows residuals $\Delta D$ as a function of reference distance, and the side histograms split aligned vs. misaligned galaxies.}
\label{app_fig:residuals}
\end{figure*}

\begin{table*}[ht]
\centering
\caption{Residual statistics for TF distance comparisons.}
\label{app_tab:method to method}
\begin{tabular}{lccc}
\hline
Parameter & Kinematic vs Tractor & Kinematic vs Hostphot & Hostphot vs Tractor \\
\hline
Mean $\Delta D$ & $-0.07 \pm 19.35$ Mpc & $0.87 \pm 19.60$ Mpc & $0.94 \pm 14.88$ Mpc \\
Median $\Delta D$ & $1.98$ Mpc & $-0.23$ Mpc & $0.23$ Mpc \\
RMS & $19.35$ Mpc & $19.62$ Mpc & $14.91$ Mpc \\
Outliers ($>3\sigma$) & $1.9\%$ & $0.9\%$ & $1.4\%$ \\
\hline
Aligned galaxies: Mean $\Delta D$ & $-0.26$ Mpc & $-0.27$ Mpc & $-0.01$ Mpc \\
Aligned galaxies: Std $\Delta D$ & $14.52$ Mpc & $14.27$ Mpc & $6.35$ Mpc \\
Misaligned galaxies: Mean $\Delta D$ & $0.41$ Mpc & $3.76$ Mpc & $3.35$ Mpc \\
Misaligned galaxies: Std $\Delta D$ & $28.07$ Mpc & $28.79$ Mpc & $25.90$ Mpc \\
\hline
\end{tabular}
\tablefoot{Columns: (1) Statistics; (2) \texttt{Kinematic} vs. \texttt{Tractor} distances; (3) \texttt{Kinematic} vs. \texttt{HostPhot} distances; (4) \texttt{HostPhot} vs. \texttt{Tractor} distances. For each method pair, the table lists the mean and median residuals ($\Delta D$), the RMS scatter, and the fraction of $>3\sigma$ outliers. The lower rows report the same statistics separately for aligned and misaligned galaxies, illustrating the substantially larger scatter associated with  
PA misalignment and morphological disturbance.}
\end{table*}

\begin{table*}
\centering
\caption{Sensitivity test for including UGC\,2319 in the ODR TF calibration.}

\label{app: sensitivity}
\begin{tabular}{lcccccc}
\hline
 & \multicolumn{2}{c}{Kinematic} 
 & \multicolumn{2}{c}{HostPhot} 
 & \multicolumn{2}{c}{Tractor} \\
\cline{2-3} \cline{4-5} \cline{6-7}
Quantity 
 & Adopted & +UGC\,2319 
 & Adopted & +UGC\,2319 
 & Adopted & +UGC\,2319 \\
\hline
Slope 
 & $-6.15 \pm 0.88$ 
 & $-54.27 \pm 64.97$ 
 & $-5.79 \pm 0.75$ 
 & $-12.42 \pm 4.81$ 
 & $-6.09 \pm 0.74$ 
 & $-12.17 \pm 4.58$ \\
Zero-point 
 & $-22.58 \pm 0.16$ 
 & $-32.70 \pm 13.28$ 
 & $-22.63 \pm 0.20$ 
 & $-24.39 \pm 1.12$ 
 & $-22.70 \pm 0.20$ 
 & $-24.40 \pm 1.15$ \\
Intrinsic scatter 
 & $0.58 \pm 0.16$ 
 & $7.85 \pm 17.28$ 
 & $0.49 \pm 0.12$ 
 & $1.71 \pm 7.19$ 
 & $0.42 \pm 0.07$ 
 & $1.47 \pm 3.47$ \\
Comment 
 & Stable fit 
 & Unstable/unphysical 
 & Stable 
 & Degraded 
 & Stable 
 & Degraded \\
\hline
\end{tabular}

\tablefoot{The ``Adopted'' calibration excludes UGC\,2319, whereas ``+UGC\,2319'' includes it. UGC\,2319 is more than two magnitudes brighter than any other calibrator and lies at the edge of the velocity distribution, giving it disproportionate leverage in the orthogonal-distance regression (ODR) fit. Its inclusion drives large shifts in slope and zero point and increases the scatter by an order of magnitude, indicating that the resulting TF relation is not physically meaningful.}
\end{table*}

\break

\section{Photometric and kinematic distance measurements }\label{app: appendix_distance}

\begin{table*}[h]
\centering
\caption{Absolute magnitudes and distances from the three PA determination methods.} \label{tab:distances}
\tiny
\begin{tabular}{l|c|c|cccc|cccc|cccc}
\toprule
\hline
Name & RA & Dec & \multicolumn{4}{c}{Kinemetry} & \multicolumn{4}{c}{HostPhot} & \multicolumn{4}{c}{Tractor} \\
     & (deg) & (deg) & $M_K$ & $\sigma_{M_K}$ & $D_K$ & $\sigma_{D_K}$ & $M_H$ & $\sigma_{M_H}$ & $D_H$ & $\sigma_{D_H}$ & $M_T$ & $\sigma_{M_T}$ & $D_T$ & $\sigma_{D_T}$ \\
     &       &       & (mag) & (mag)          & (Mpc) & (Mpc)          &  &           &  &         &  &          &  &          \\
\midrule
UGC07012       & 180.51 & 29.85 & -19.61 & 0.21 & 47.57 & 4.59 & -19.66 & 0.14 & 48.72 & 3.11 & -19.64 & 0.16 & 48.28 & 3.47 \\
MCG-02-03-015  & 11.94  & -9.84 & -20.79 & 0.20 & 53.93 & 4.98 & -21.12 & 0.21 & 62.76 & 6.03 & -20.99 & 0.10 & 58.96 & 2.80 \\
UGC08988       & 210.83 & 60.99 & -19.76 & 1.11 & 63.40 & 32.52 & -19.80 & 1.20 & 64.61 & 35.66 & -19.39 & 0.59 & 53.49 & 14.54 \\
UGC08004       & 192.91 & 31.35 & -20.81 & 0.29 & 93.46 & 12.39 & -20.66 & 0.25 & 87.13 & 10.20 & -20.67 & 0.28 & 87.46 & 11.48 \\
NGC2526        & 121.74 & 8.00  & -21.11 & 0.65 & 75.87 & 22.75 & -20.95 & 0.63 & 70.65 & 20.36 & -20.92 & 0.66 & 69.67 & 21.10 \\
UGC10331       & 244.34 & 59.32 & -18.24 & 0.91 & 25.93 & 10.83 & -19.48 & 0.20 & 45.78 & 4.29 & -19.40 & 0.21 & 44.21 & 4.31 \\
NGC5559        & 214.80 & 24.80 & -21.30 & 0.19 & 90.97 & 8.02 & -21.30 & 0.19 & 90.69 & 8.02 & -21.42 & 0.19 & 96.11 & 8.35 \\
NGC0171        & 9.34   & -19.93& -19.73 & 0.52 & 21.76 & 5.17 & -21.13 & 0.58 & 41.50 & 11.06 & -20.03 & 0.53 & 25.02 & 6.14 \\
UGC00139       & 3.63   & -0.74 & -20.61 & 0.38 & 62.49 & 11.02 & -20.55 & 0.50 & 60.77 & 14.05 & -20.75 & 0.32 & 66.50 & 9.84 \\
UGC01749       & 34.06  & 18.31 & -21.60 & 0.09 & 125.77& 5.65 & -21.71 & 0.09 & 132.41& 5.62 & -21.73 & 0.11 & 133.54& 6.65 \\
NGC0234        & 10.88  & 14.34 & -22.87 & 0.16 & 109.12& 8.03 & -21.48 & 0.17 & 57.62 & 4.43 & -21.45 & 0.19 & 56.80 & 5.01 \\
NGC5622        & 216.55 & 48.56 & -21.00 & 0.26 & 66.49 & 7.91 & -21.20 & 0.21 & 73.09 & 7.02 & -21.18 & 0.22 & 72.20 & 7.30 \\
NGC3687        & 172.00 & 29.51 & -21.04 & 0.61 & 47.85 & 13.52 & -21.08 & 0.61 & 48.84 & 13.79 & -20.70 & 0.39 & 41.04 & 7.40 \\
NGC7536        & 348.55 & 13.43 & -20.82 & 0.15 & 59.76 & 4.11 & -21.01 & 0.13 & 65.41 & 3.93 & -21.02 & 0.14 & 65.56 & 4.19 \\
UGC08107       & 194.92 & 53.34 & -22.20 & 0.71 & 109.24& 35.86 & -22.44 & 0.69 & 122.41& 38.76 & -22.34 & 0.71 & 116.64& 37.91 \\
\bottomrule
\end{tabular}

\tablefoot{Columns: (1) Galaxy name; (2) Right Ascension (RA, J2000 degrees); (3) Declination (Dec, J2000 degrees); (4–7) Kinemetry absolute magnitude ($M_K$) and distance ($D_K$) with uncertainties; (8–11) \texttt{HostPhot} absolute magnitude  
($M_H$) and distance ($D_H$) with uncertainties; (12–15) \texttt{tractor} absolute magnitude ($M_T$) and distance ($D_T$) with uncertainties.}
\end{table*}

\begin{table*}[h!]
\centering
\caption{Comparison between HostPhot and \texttt{Tractor} photometric orientation measurements.}
\label{app_tab: Photometric}
\begin{tabular}{lccccccc}
\hline
Name & Ra & Dec & R26 & Host\_ang & Track\_ang & $\Delta$PA & Outlier \\
\hline
IC4582            & 236.414375  & 28.0886666  & 71.00332642  & 81.3491103  & 81.8   & -0.450889702  & inlier \\
ASASSN14hr        & 27.671922   & -14.517582  & 34.49731827  & 23.25604622 & 16.2   & 7.05604622    & inlier \\
ASASSN14co        & 239.376176  & 1.112121    & 26.59677505  & 103.9223614 & 105.8  & -1.877638598  & inlier \\
UGC11241          & 276.7075833 & 51.1391666  & 44.18946075  & 118.8576765 & 129    & -10.14232349  & outlier \\
iPTF14ans         & 226.123966  & 2.333006    & 39.20230103  & 106.9758453 & 123.3  & -16.32415472  & outlier \\
PGC10254          & 40.620875   & 18.1647222  & 41.60279846  & 123.0760243 & 121.1  & 1.976024339   & inlier \\
UGC09842          & 231.2741667 & 37.9601666  & 60.53148651  & 164.5696669 & 165.2  & -0.630333071  & inlier \\
UGC04233          & 121.8219583 & 8.86913888  & 61.65338898  & 23.60837811 & 23.8   & -0.191621889  & inlier \\
UGC09537          & 222.1114583 & 34.9980555  & 113.323288   & 49.35564362 & 47.7   & 1.655643622   & inlier \\
UGC10337          & 244.8691667 & 7.27875     & 73.89341736  & 154.9906391 & 154.1  & 0.890639099   & inlier \\
UGC10811          & 259.6821667 & 58.1351111  & 70.18004608  & 0.665120993 & 5.8    & -5.134879007  & inlier \\
NGC5888           & 228.2807083 & 41.2646388  & 58.47261429  & 62.8423644  & 70.6   & -7.757635598  & inlier \\
MCG-01-09-006     & 47.20133333 & -7.0406111  & 89.692276    & 104.8451063 & 102    & 2.845106298   & inlier \\
NGC5837           & 226.1690833 & 12.6334444  & 49.63430786  & 115.4208824 & 103    & 12.42088242   & outlier \\
UGC11758          & 322.7400833 & 13.9860833  & 81.36412048  & 148.9388746 & 148.8  & 0.138874636   & inlier \\
\hline
\end{tabular}
\tablefoot{Columns: (1) Galaxy name; (2) Right Ascension (RA, J2000 degrees); (3) Declination (Dec, J2000 degrees); (4) Photometric radius $R_{26}$ (arcsec); (5) HostPhot position angle (PA); (6) \texttt{Tractor} position angle (PA); (7) PA difference $\Delta{\rm PA}$; (8) Outlier flag indicating whether the galaxy exceeds the outlier threshold ($|\Delta{\rm PA}| > 10^\circ$).}
\end{table*}

\clearpage

\section{Kinemetry analysis of IC\,1199} 

\begin{figure*}[h]
\centering
\includegraphics[width=0.32\textwidth]{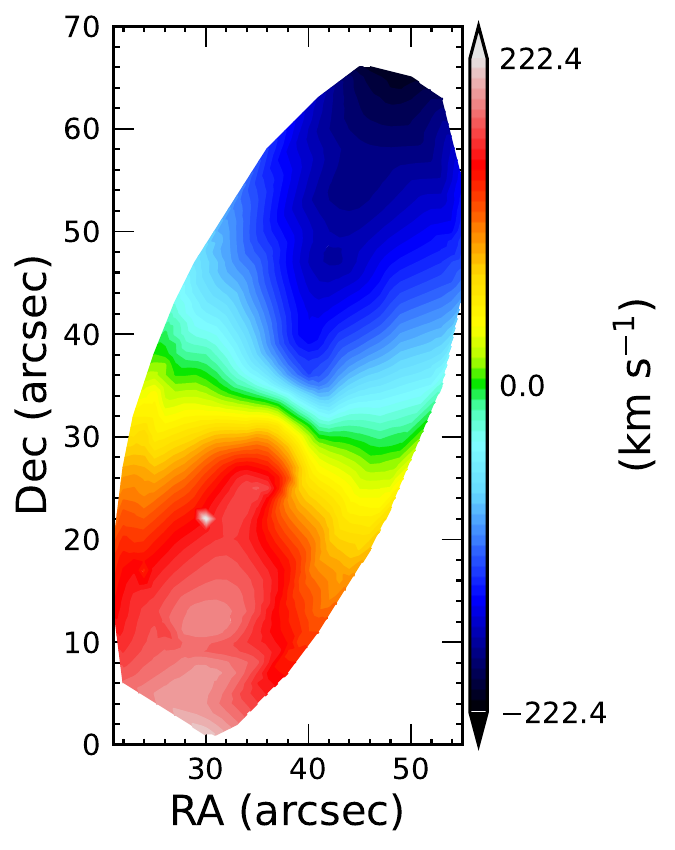}
\hfill
\includegraphics[width=0.32\textwidth]{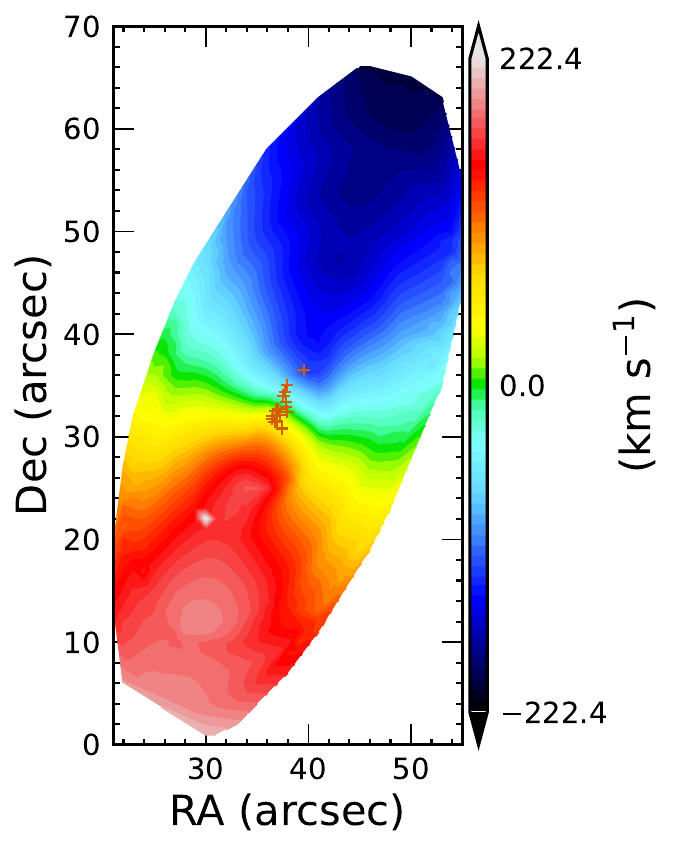}
\hfill
\includegraphics[width=0.32\textwidth]{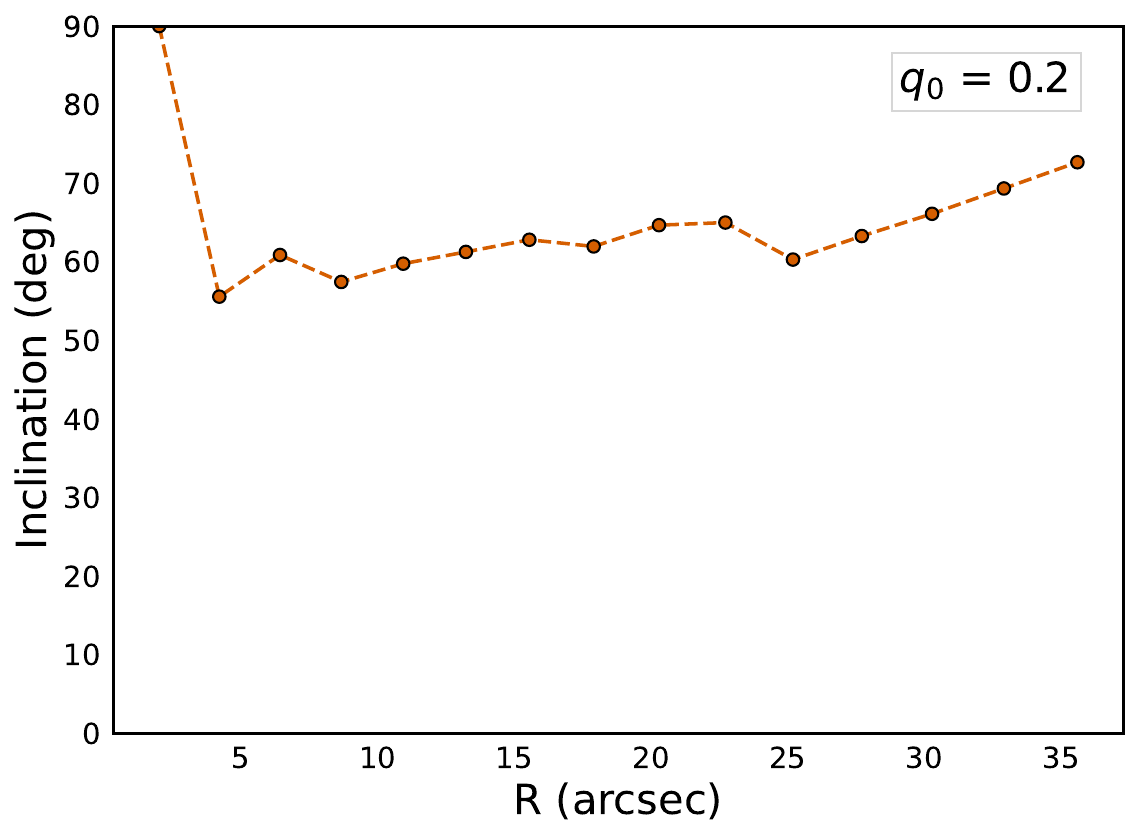}

\vspace{0.3cm}

\makebox[0.32\textwidth][c]{\small (a) Observed velocity field}
\hfill
\makebox[0.32\textwidth][c]{\small (b) Model velocity field}
\hfill
\makebox[0.32\textwidth][c]{\small (c) Inclination angle}

\vspace{0.6cm}

\includegraphics[width=0.32\textwidth]{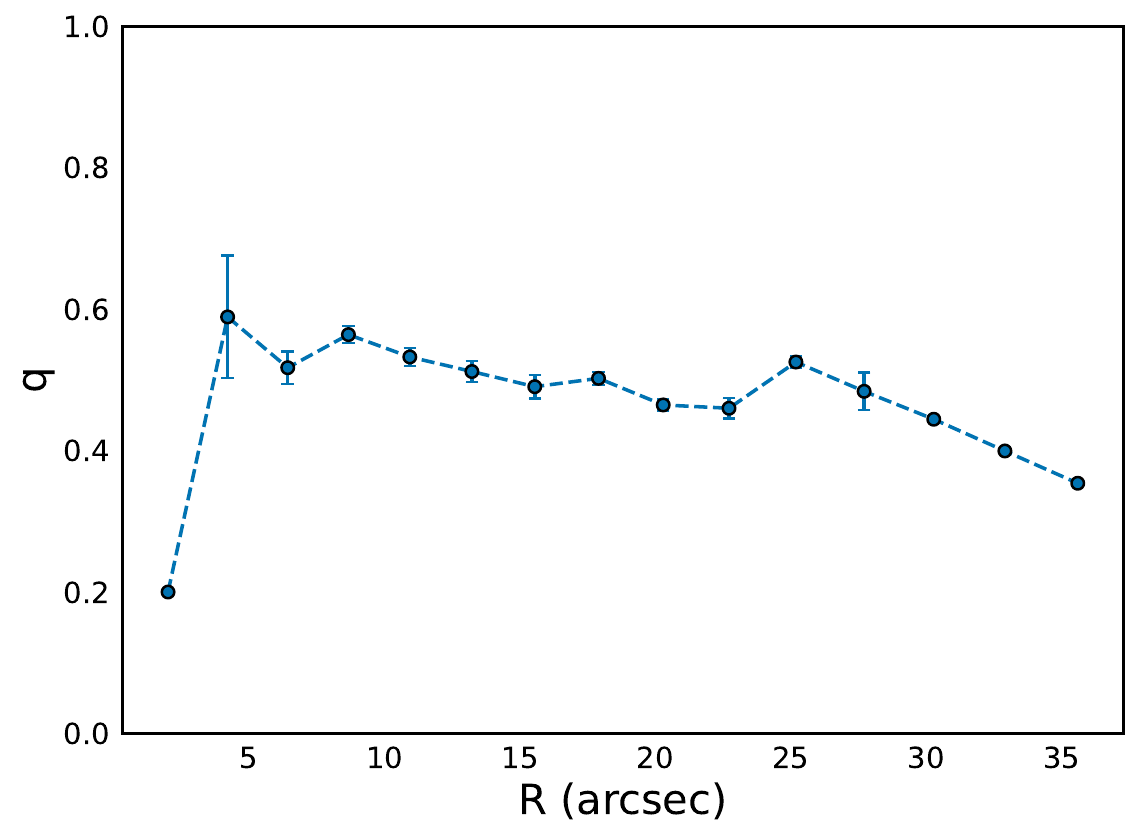}
\hfill
\includegraphics[width=0.32\textwidth]{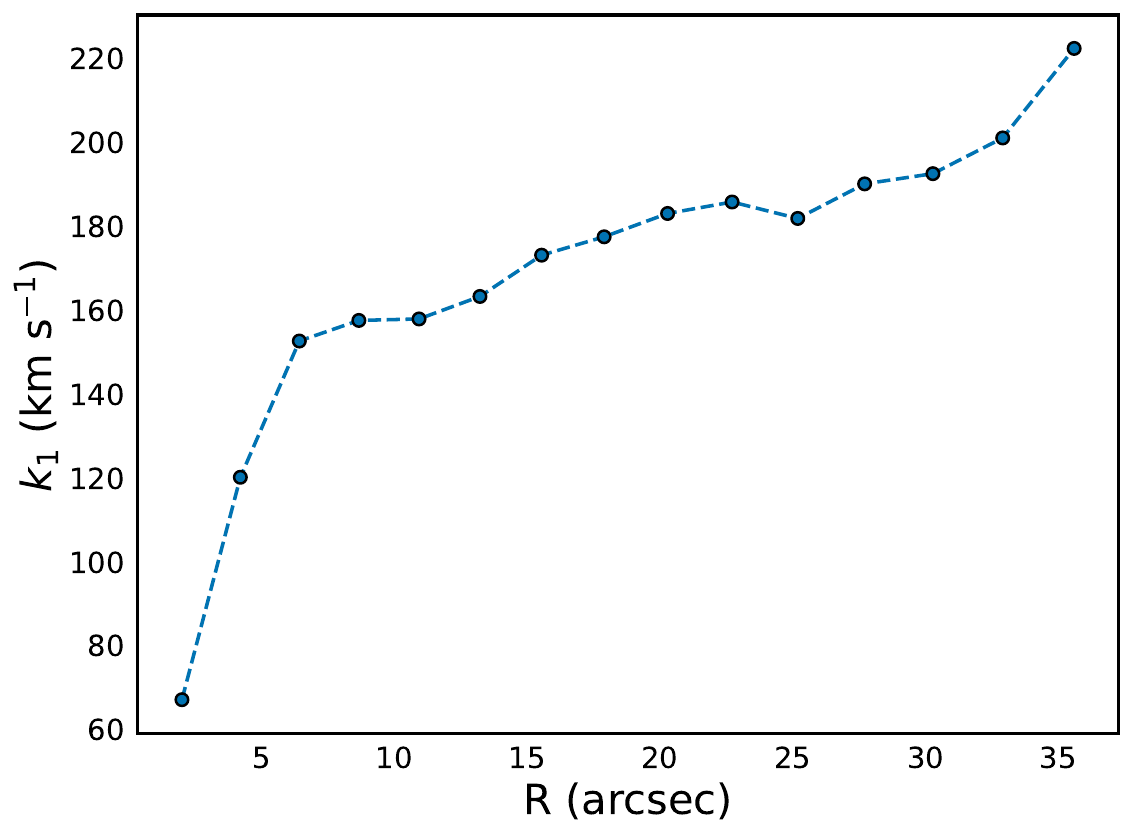}
\hfill
\includegraphics[width=0.32\textwidth]{figure/IC1199_velocity_profile_PA.pdf}

\vspace{0.3cm}

\makebox[0.32\textwidth][c]{\small (d) Flattening ($q$)}
\hfill
\makebox[0.32\textwidth][c]{\small (e) Radial velocity ($k_1$)}
\hfill
\makebox[0.32\textwidth][c]{\small (f) PA}

\vspace{0.6cm}

\includegraphics[width=0.32\textwidth]{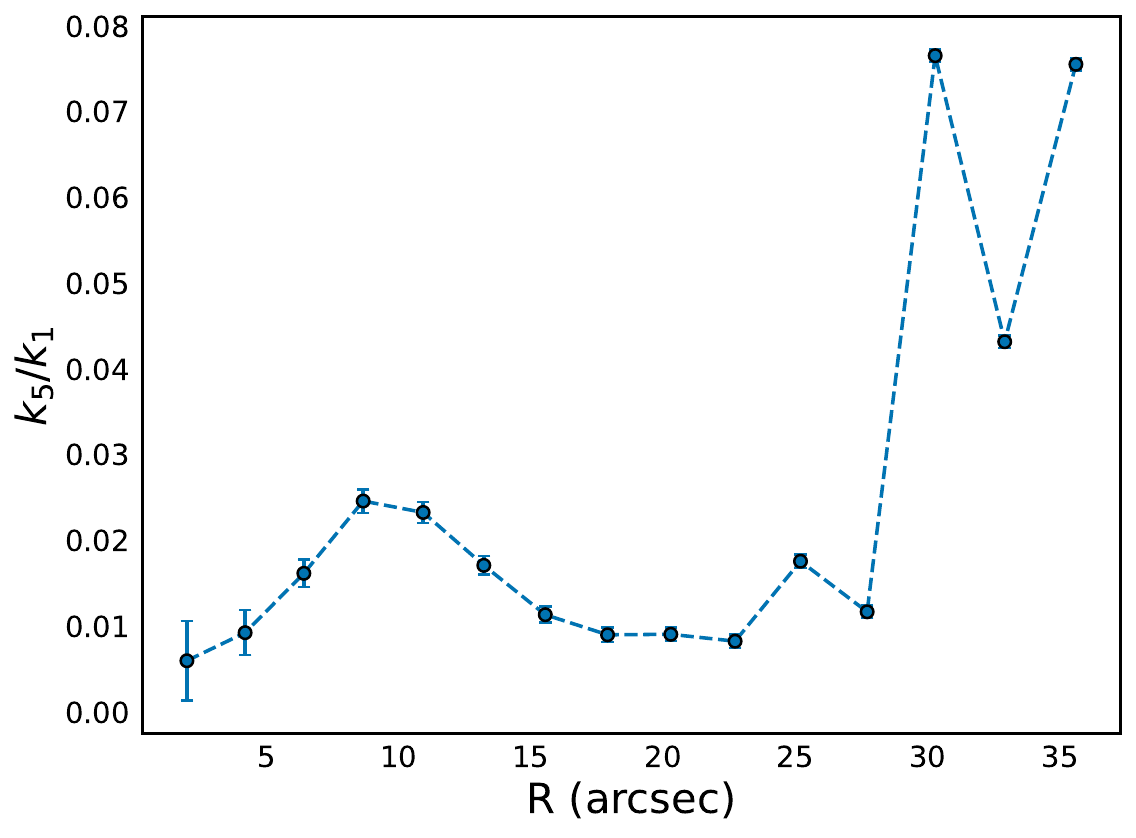}
\hfill
\includegraphics[width=0.32\textwidth]{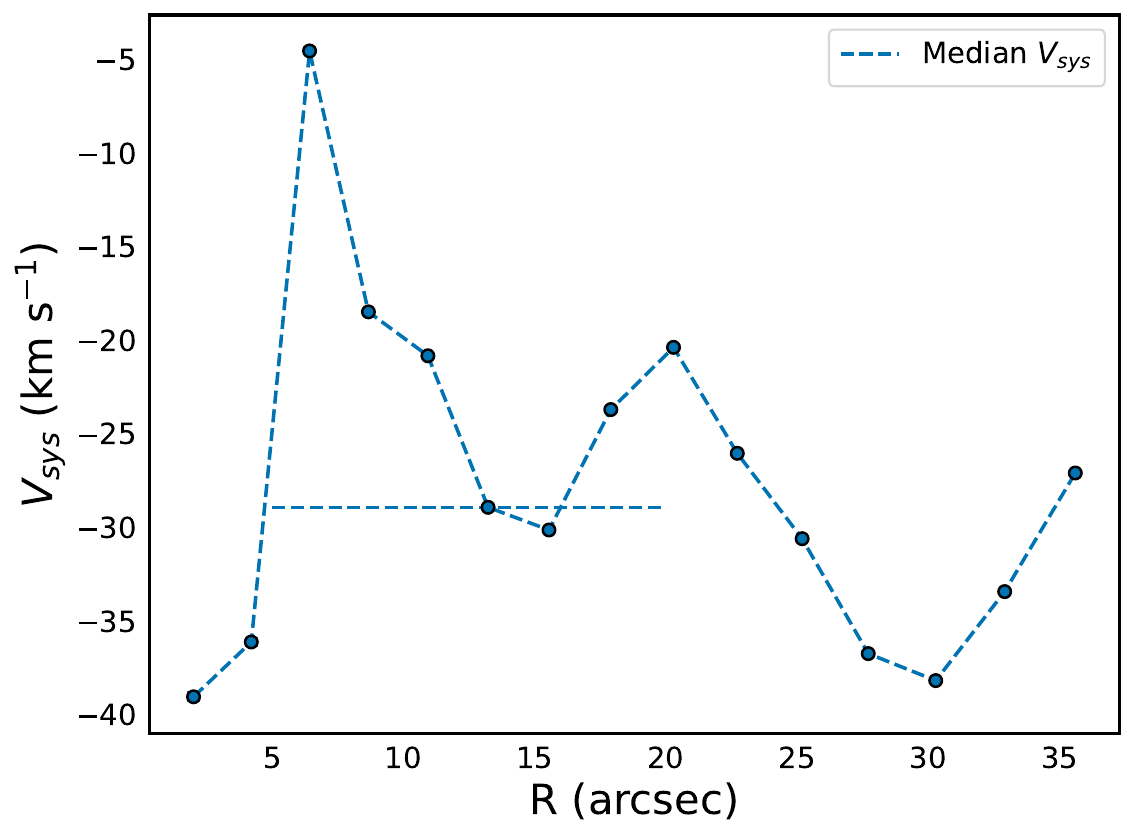}
\hfill
\includegraphics[width=0.32\textwidth]{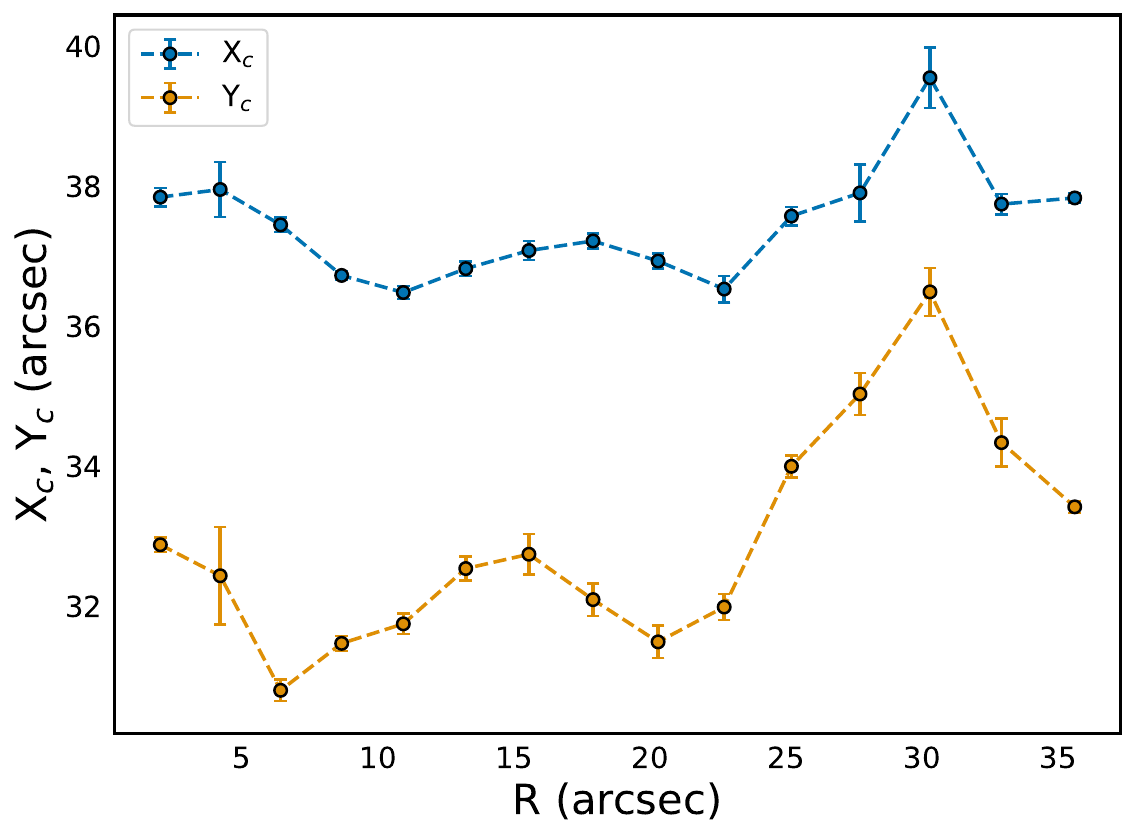}

\vspace{0.3cm}

\makebox[0.32\textwidth][c]{\small (g) Higher-order term ($k_{5,1}$)}
\hfill
\makebox[0.32\textwidth][c]{\small (h) Systemic velocity ($V_{\mathrm{sys}}$)}
\hfill
\makebox[0.32\textwidth][c]{\small (i) Kinematic center ($X_c$, $Y_c$)}

\vspace{0.5cm}

\caption{Comprehensive kinemetry analysis of IC\,1199. Top row: (a) Observed, (b) modeled, and (c) Inclination angle. Middle row: (d) Axis ratio ($q$), (e) radial velocity amplitude ($k_1$), and (f) kinematic PA. Bottom row: (g) Higher-order harmonic ($k_{5,1}$), (h) systemic velocity ($V_{\mathrm{sys}}$), and (i) fitted kinematic center ($X_c$, $Y_c$). Gray bands represent 1$\sigma$ uncertainties; dashed lines indicate photometric priors.}
\label{app_sec:kinemetry}
\end{figure*}

\clearpage
\section{Aligned anomaly} \label{App: aligned_anamoly}

\begin{figure*}[h]
\centering
\includegraphics[width=0.48\textwidth]{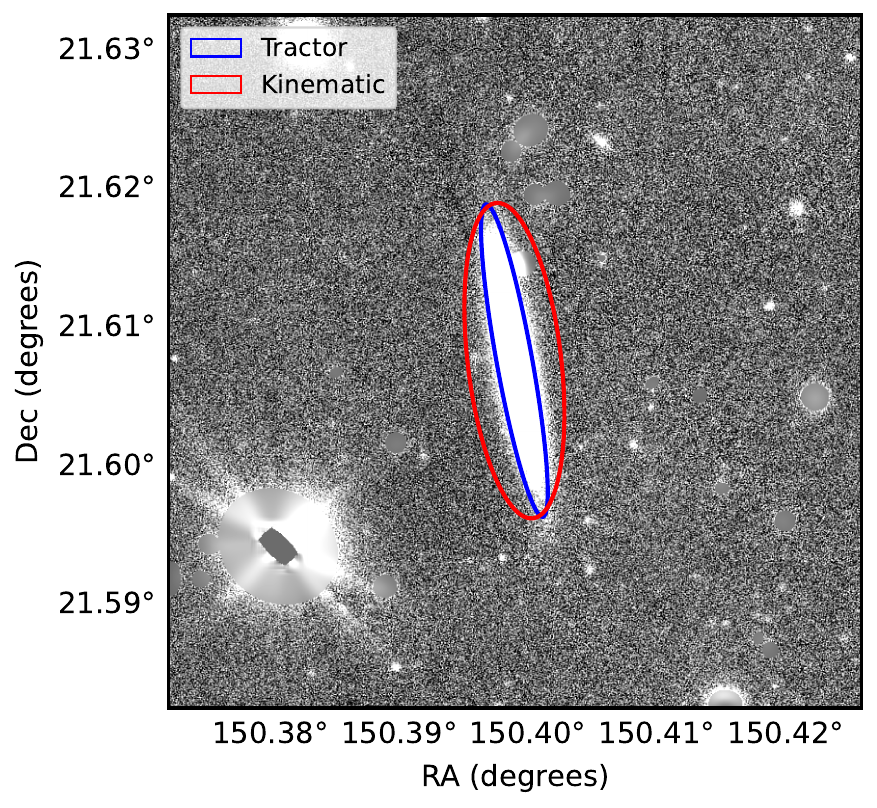}
\includegraphics[width=0.48\textwidth]{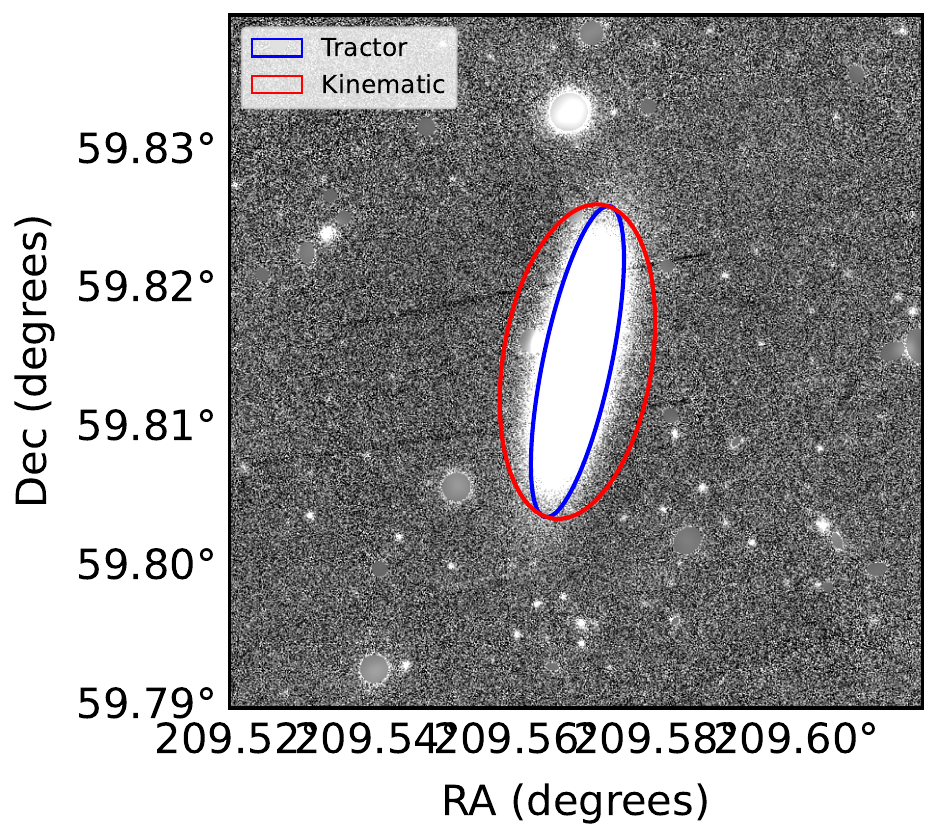}
\caption{Projected ellipses overlaid on the optical image, showing \texttt{`tractor'} (blue), and \texttt{`kinematic'} (red) geometries. All ellipses are sky-plane projections. In highly inclined systems, small differences in PA translate into large uncertainties in inferred rotational velocities and thus distance.}
\label{app_fig:ellipse}
\end{figure*}

\end{appendix}

\end{document}